\documentclass[sn-standardnature,iicol]{sn-jnl}

\usepackage{siunitx}
\usepackage{amssymb}
\usepackage{times}
\usepackage{graphicx}
\usepackage{makecell}
\usepackage{rotating}
\usepackage{multibib} 
\usepackage{anyfontsize}
\usepackage{comment}
\usepackage{placeins}
\usepackage{ragged2e}
\newcites{App}{Methods References}

\newcommand{\araa}{Annu.\ Rev.\ Astron.\ Astrophys.} 
\newcommand{\aj}{Astron.\ J.} 
\newcommand{\apj}{Astrophys.\ J.} 
\newcommand{\apjl}{Astrophys.\ J.\ Lett.} 
\newcommand{\apjs}{Astrophys.\ J.\ Suppl.\ Ser.} 
\newcommand{\aap}{Astron.\ Astrophys.} 
\newcommand{\icarus}{Icarus} 
\newcommand{\jqsrt}{J.\ Quant.\ Spectrosc.\ Radiat.\ Transf.} 
\newcommand{\mnras}{Mon.\ Not.\ R.\ Astron.\ Soc.} 
\newcommand{\nat}{Nature} 
\newcommand{\natas}{Nat.\ Astron.} 
\newcommand{\pasp}{Publ.\ Astron.\ Soc.\ Pac.} 
\newcommand{\ssr}{Space\ Sci.\ Rev.} 

\usepackage{soul}

\begin{document}

\title{\centering{Challenges in the detection of gases in exoplanet atmospheres}}

\author*[1]{\fnm{Luis} \sur{Welbanks}}\equalcont{}\email{luis.welbanks@asu.edu }
\author[1,2]{\fnm{Matthew C.} \sur{Nixon}}\equalcont{}
\author[3]{\fnm{Peter} \sur{McGill}}
\author[1]{\fnm{Lana J.} \sur{Tilke}}
\author[1]{\fnm{Lindsey S.} \sur{Wiser}}
\author[1]{\fnm{Yoav} \sur{Rotman}}
\author[1,4,5]{\fnm{Sagnick} \sur{Mukherjee}}
\author[6]{\fnm{Adina D.} \sur{Feinstein}}
\author[1]{\fnm{Michael R.} \sur{Line}}
\author[7,8]{Bj\"orn Benneke}
\author[9,10,11]{\fnm{Sara} \sur{Seager}}
\author[12]{\fnm{Thomas G.} \sur{Beatty}}
\author[6]{\fnm{Darryl Z.} \sur{Seligman}}
\author[13]{\fnm{Vivien} \sur{Parmentier}}
\author[5]{\fnm{David K.} \sur{Sing}}


\affil[1]{\orgdiv{School of Earth and Space Exploration}, \orgname{Arizona State University}, \orgaddress{\city{Tempe}, \state{AZ}, \country{USA}}}
\affil[2]{\orgdiv{Department of Astronomy}, \orgname{University of Maryland}, \orgaddress{\city{College Park}, \state{MD}, \country{USA}}}
\affil[3]{\orgdiv{Space Science Institute}, \orgname{Lawrence Livermore National Laboratory}, \orgaddress{\city{Livermore}, \state{CA}, \country{USA}}}
\affil[4]{\orgdiv{Department of Astronomy and Astrophysics}, \orgname{University of California Santa Cruz}, \orgaddress{\city{Santa Cruz}, \state{CA}, \country{USA}}}
\affil[5]{\orgdiv{William H.\ Miller III Department of Physics and Astronomy}, \orgname{Johns Hopkins University}, \orgaddress{\city{Baltimore}, \state{MD}, \country{USA}}}
\affil[6]{\orgdiv{Department of Physics and Astronomy}, \orgname{Michigan State University}, \orgaddress{\city{East Lansing}, \state{MI}, \country{USA}}}
\affil[7]{Department of Earth, Planetary, and Space Sciences, University of California, Los Angeles, CA USA}
\affil[8]{Department of Physics and Trottier Institute for Research on Exoplanets, Université de Montréal, Montreal, QC, Canada}
\affil[9]{\orgdiv{Department of Earth, Atmospheric and Planetary Sciences}, \orgname{Massachusetts Institute of Technology}, \orgaddress{\city{Cambridge}, \state{MA}, \country{USA}}}
\affil[10]{\orgdiv{Department of Physics}, \orgname{Massachusetts Institute of Technology}, \orgaddress{\city{Cambridge}, \state{MA}, \country{USA}}}
\affil[11]{\orgdiv{Department of Aeronautical and Astronautical Engineering}, \orgname{Massachusetts Institute of Technology}, \orgaddress{\city{Cambridge}, \state{MA}, \country{USA}}}
\affil[12]{\orgdiv{Department of Astronomy}, \orgname{University of Wisconsin-Madison}, \orgaddress{\city{Madison}, \state{WI}, \country{USA}}}
\affil[13]{\orgdiv{Laboratoire Lagrange}, \orgname{Observatoire de la Côte d’Azur, Université Côte d’Azur}, \orgaddress{\city{Nice}, \country{France}}}

\abstract{Claims of detections of gases in exoplanet atmospheres often rely on comparisons between models including and excluding specific chemical species. However, the space of molecular combinations available for model construction is vast and highly degenerate.  Only a limited subset of these combinations is typically explored for any given detection. As a result, apparent detections of trace gases risk being artifacts of incomplete modeling rather than robust identification of atmospheric constituents, especially in the low signal-to-noise regime. Using the sub-Neptune K2-18 b as a case study, we show that recent biosignature claims vanish when the model space is expanded, with numerous alternatives providing equally good or better fits. We demonstrate that the significance of a claimed detection relies on the choice of models being compared, and that model preference does not in itself imply the presence of a specific gas. We recommend treating model comparisons instead as relative adequacy tests, which should be supported by theoretical predictions and complementary metrics of statistical significance in order to attribute a signal to a particular gas.}

\maketitle

Remote sensing of exoplanet atmospheres hinges on our ability to reliably interpret signatures of absorption, emission, and scattering by atmospheric chemical species and to infer their abundances. The high precision and broad wavelength coverage of the James Webb Space Telescope (JWST)\cite{Gardner2006} enables atmospheric characterization of sub-Neptune and terrestrial exoplanets. Even with JWST, spectra of these objects often operate near the limits of achievable signal-to-noise ratios. In the absence of strong, statistically significant features, the community frequently couples atmospheric models with statistical inference methods (commonly called \textit{atmospheric retrievals}) to tease out small signals in the data and draw conclusions about the nature of these distant worlds.

While the atmospheric retrieval approach is generally regarded as agnostic and data-driven, it still has the potential to mislead regarding the strength of an apparent detection, and the subsequent interpretation of planetary conditions. The field has adopted Bayesian model comparison as a standard method to determine the presence of chemical species and physical processes in exoplanet atmospheres\cite{Benneke2013,Sedaghati2017,Fu2024}. 
However, concerns have emerged that relying on the Bayesian evidence as the sole goodness-of-fit metric is insufficient for robust atmospheric interpretation\cite{Wilson2021}. Recent work has advocated for methods beyond the Bayesian evidence that provide insight into which parts of a dataset drive inferences\cite{Vehtari2017, Welbanks2023}, and for averaging inferences over models instead of selecting between them\cite{Nixon2024a}. However, the pitfalls of Bayesian model comparison in regimes lacking strong spectral features that can be clearly identified, attributed, and interpreted (see Key Criteria from ref.\cite{Seager2025}) remain largely unexplored.

\begin{figure*}[h!]
    \centering
    \includegraphics[width=\linewidth]{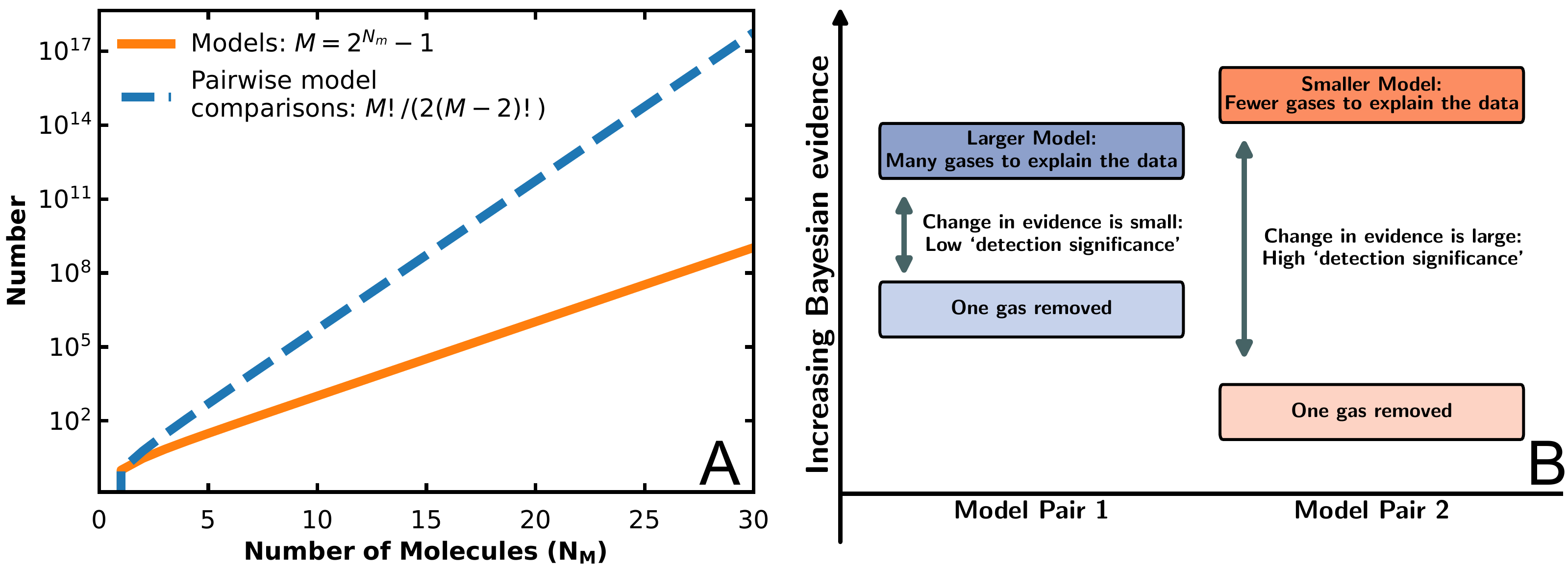}
    \caption{\textbf{Combinatorial growth of model space and dependence of apparent detection significance on reference model.} \textbf{A)} 
    Number of unique models and pairwise model comparisons that would be needed to exhaustively search the model space as a function of the number of molecules considered. Exploring any significant fraction of this space for typical models with $>10$ molecules is computationally infeasible. These numbers are lower limits because they do not account for other discrete atmospheric model choices such as cloud and pressure-temperature profile parameterizations.  \textbf{B)} Schematic showing how the apparent detection significance from Bayesian model comparison depends on the chosen reference model.}
    \label{fig:model_space}
\end{figure*}

Here we demonstrate that Bayesian model comparison does not always translate into robust detections of atmospheric constituents. Pairwise model comparisons  only quantify the relative fit to the data of one model against another. When all considered candidate models are poor representations of reality, the best-performing model is simply the least inadequate and may not necessarily lead to reliable interpretations of the data. Conversely, when all candidate models adequately fit a spectrum, a preference for one model over another does not rule out the worse-performing model. We demonstrate how these issues manifest in a case study of the sub-Neptune K2-18~b\cite{Benneke2019,foreman-mackey2015}, where the inferred atmospheric properties can be shaped as much by preconceived notions for what this planet ought to be like as by the observational data.

\paragraph{Bayesian Model Comparison in Exoplanet Atmospheric Inference}

Determining appropriate, data-driven model assumptions was an early challenge in analyzing exoplanet spectra given our limited \textit{a priori} knowledge of these atmospheres. Bayesian model comparison was introduced to compare competing hypotheses in exoplanetary atmospheric studies\cite{Benneke2013}. Ref.\cite{Benneke2013} introduced an explicit methodology of computing the Bayesian evidence for one retrieval model that should cover the full prior hypothesis space, and comparing it to models for which one chemical species (or type of aerosol) was removed at a time from that otherwise full prior hypothesis space. Although exploring a complete prior hypothesis space may not be possible, as explained below, this remove-one-at-a-time approach was intended to ensure that a high confidence in the presence of a particular atmospheric constituent was reported only if no other reasonable constituent could have plausibly resulted in the observed data (Section 2.2.2 in ref.\cite{Benneke2013}). This approach was quickly adopted by the field, and became popular for its promise of applying ``Occam's razor'' to penalize the fit of more complex models over simpler ones that were not sufficiently supported by the data. 

Given the practical impossibility of constructing a full prior hypothesis space, and no well-established conventions in the field, there is a risk of Bayesian model comparison being used to evaluate the relative preference between models even when neither adequately represents the full prior hypothesis space. Detections could then be reported based on Bayes factors between models that each include only a limited subset of chemical species, without comparison or contextualization relative to a more comprehensive hypothesis space. If two similarly incomplete or inadequate models are compared, there is a risk of incorrectly interpreting a model preference as a detection. The problem worsens if this model preference is converted into a ``detection significance,'' expressed in terms of $N_\sigma$\cite{Trotta2008a}, losing the relative nature of the metric and further contributing to misinterpretation.

Use of model comparisons in isolation, without acknowledging their limitations, contributes to an opaque definition of ``detection''. Any detection significance is computed relative to a chosen reference model and therefore strongly depends on the included species and parameters. Statements of model preference without qualification of the reference model are arbitrary. Furthermore, because the number of plausible chemical species is large --- hundreds are available via the HITRAN database\cite{Sharpe2004, Gordon2022} for example --- meaningfully exploring any significant fraction of the hypothesis model space is computationally unfeasible. This practical limitation is an invitation to recognize the subjective nature in building an appropriate model space. Figure~\ref{fig:model_space} shows how the number of unique models and pairwise model comparisons needed to exhaustively explore the detection space grows with the number of molecules considered. Even for a modest set of $10$ species, the number of unique compositional models for an atmosphere is $1,024$, and the number of possible pairwise model comparisons is $523,776$. While a complete exploration of model space is not currently practical, any detection claimed through pairwise model comparison should at minimum be contextualized within the broader set of plausible comparisons that could have yielded equal or greater significance. Credible detection claims demand systematic elimination of credible alternatives.

\paragraph{Model Comparison Alone Is Not Detection}\label{sec:model_is_not_detection}

The concept of \textit{detection} in astronomy typically refers to the measurement of a signal above background noise, arising from detector electronics, intrinsic fluctuations in the photon flux, or other sources. Assuming normally distributed noise, the signal amplitude can be expressed in units of the noise's standard deviation.  A 3-$\sigma$ detection, for example, implies that the observed signal exceeds the noise by three times its standard deviation. This is a fundamentally different measure than the Bayesian model comparisons explained above. However, the shared use of $N_\sigma$-confidence has resulted in ambiguities in the meaning of detection, and in some cases telescope resources are optimized to yield $N_\sigma$ detections without clearly distinguishing whether the value is derived from a signal-to-noise argument or a model comparison.

In high signal-to-noise regimes, atmospheric detections rely on strong, well-resolved absorption/emission features spanning multiple spectral bands\cite{ERS2023,Welbanks2024,Bell2023c}. A clear example is the identification of CO$_2$ in the atmosphere of WASP-39~b\cite{ERS2023}, where an absorption feature at 4.3$\mu$m appears at $>10\sigma$ significance, referring to the signal-to-noise ratio of the data; far stronger than previously predicted\cite{Bean2018}. Attribution to CO$_2$ is supported by atmospheric models for warm gas giants and expectations from thermochemical equilibrium\cite{Lodders2002,Moses2013a}, and reinforced by the presence of additional CO$_2$ features at other wavelengths, and across independent data reductions\cite{Feinstein2023,Carter2024}. Bayesian model comparisons confirm that models including CO$_2$ best explain the observed features, but in this context, the comparison does not constitute the detection itself; it merely affirms the attribution.

\begin{figure}
    \centering
    \includegraphics[width=\linewidth]{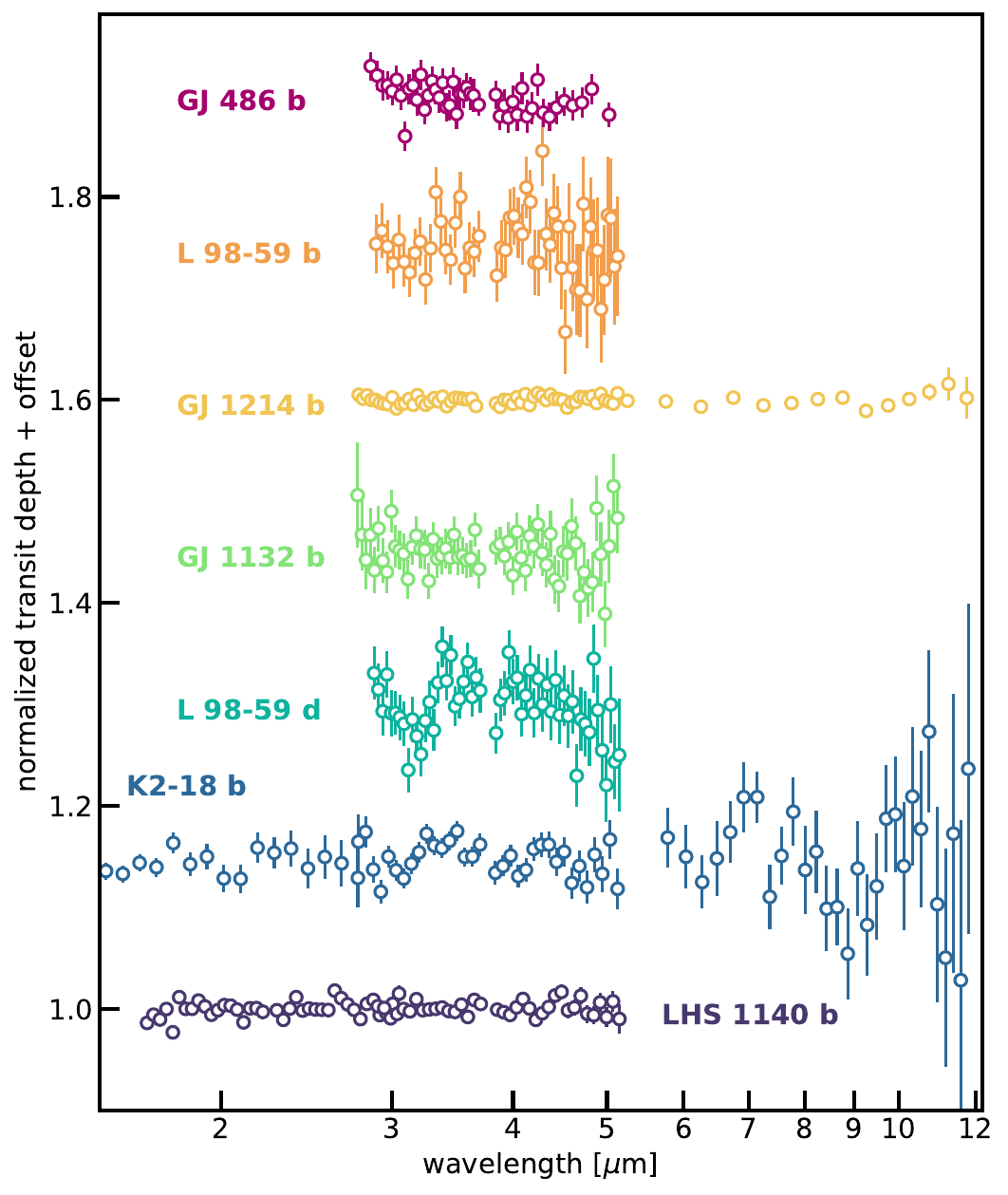}
    \caption{\textbf{Selected JWST transmission spectra of small exoplanets ($R_p<3R_{\oplus}$).} Transmission spectra for seven exoplanets --- GJ 486 b\cite{Moran2023}, L 98-59 b\cite{Bello-Arufe2025}, GJ 1214 b\cite{Schlawin2024,Kempton2023b}, GJ 1132 b\cite{May2023}, L 98-59 d\cite{Gressier2024}, K2-18 b\cite{Madhusudhan2023,Madhusudhan2025}, and LHS 1140 b\cite{Damiano2024}---observed with JWST. Colors distinguish each planet. Offsets in normalized transit depth are applied for clarity. Vertical error bars show 1$\sigma$ uncertainties (standard deviations) on the measured transit depth; data points without visible bars have smaller errors than the plotted symbols. These spectra have been analysed using Bayesian model comparison to determine preference for an atmospheric model over a featureless spectrum.}
    \label{fig:flat_family}
\end{figure}

While CO$_2$ in WASP-39~b demonstrates a clear detection grounded in strong data and physical plausibility, many recent JWST spectra of smaller exoplanets operate near the limits of low signal-to-noise, where interpretation becomes more ambiguous. JWST has delivered spectra of some of the smallest and coldest exoplanets studied to date\cite{Moran2023,Bello-Arufe2025,Schlawin2024,Kempton2023b,May2023,Gressier2024,Madhusudhan2023,Madhusudhan2025,Damiano2024}, yet several of these observations remain statistically consistent with a featureless spectrum (see Figure~\ref{fig:flat_family}). Testing whether a spectrum is consistent with an absence of features is often conducted using a frequentist approach, which tests whether a null hypothesis can be rejected at a given confidence level (see Methods and Extended Data Figure \ref{fig:ED1}). This contrasts with Bayesian model comparison, which quantifies the relative preference between two models. 

An example of this distinction is the atmospheric characterization of the sub-Earth-sized planet L~98-59~b\cite{Demangeon2021}, observed with JWST/NIRSpec\cite{BelloArufe2025}, Figure~\ref{fig:flat_family}. Prior to these observations, theoretical work\cite{Seligman2024} had identified L~98-59~b as a candidate for detecting extrasolar volcanism, predicting that SO$_2$ absorption features could emerge if tidal heating rates resembled those of Io\cite{Goldreich1966,Fortin2022}. The JWST data were found to be statistically consistent with a featureless spectrum based on $\chi^2$ analysis\cite{BelloArufe2025}. Nonetheless, Bayesian model comparisons favored atmospheric models over a bare-rock model at $\sim3.6\sigma$ significance, and models including SO$_2$ over models without it at $\sim2.4\sigma$\cite{BelloArufe2025}.  Critically, the study acknowledged the limitations of this framework, emphasizing that model comparisons are conditional and that a no-atmosphere model could not be ruled out. Confirming the existence of volcanic activity in an exoplanet would require complementary lines of evidence beyond the presence of SO$_2$ in its atmosphere, such as measurements of other chemical products resulting from volcanic activity\cite{Spencer2005} or the detection of a circumstellar plasma torus\cite{Kislyakova2018}. 

Degeneracies between atmospheric parameters further contribute to the complexity of exoplanet atmospheric retrieval. Numerous molecules have overlapping absorption bands, meaning that several chemical species can often explain a given feature in an observed spectrum. An example of this degeneracy appears in the analysis of the $1.1$--$1.8~\mu$m HST/WFC3 transmission spectrum of the sub-Neptune K2-18~b\cite{Benneke2019}. The spectrum exhibits a prominent feature near $1.4~\mu$m, initially attributed to H$_2$O at $3.9\sigma$\cite{Benneke2019}, $3.6\sigma$\cite{Tsiaras2019}, and $3.3\sigma$\cite{Welbanks2019b, Madhusudhan2020} significance by four independent analyses, all using Bayesian model comparisons. Interpreted as a frequentist significance, this would imply a $>$99.7\% probability that H$_2$O is present.  However, a subsequent study demonstrated that model realizations with methane as the dominant absorber instead of water also provide adequate fits to the data ($\chi^2/N_{\rm data} < 1.07$)\cite{Bezard2022}. 

Follow-up observations with JWST/NIRISS and NIRSpec, spanning a broader spectral range (0.9--5.2$\mu$m), revealed a preference for methane at $4$--$5\sigma$ significance, depending on model assumptions, with no strong evidence for H$_2$O\cite{Madhusudhan2023, Schmidt2025}. The extended coverage from JWST helped resolve the degeneracy between H$_2$O and CH$_4$ at HST/WFC3 wavelengths. Comparing the HST and JWST observations, it appears that some previous attributions to H$_2$O may have been influenced by two data points near the blue edge of the WFC3 bandpass (see Figure 6 in ref.\cite{Madhusudhan2023}). Furthermore, one earlier study did not consider CH$_4$ absorption in its atmospheric models\cite{Tsiaras2019}. The case of the HST/WFC3 spectrum of K2-18~b highlights the risks of interpreting model preference as a definitive chemical detection---particularly when the data are limited in spectral coverage and models are highly simplified.

\begin{figure*}[!htbp]
     \centering
     \includegraphics[width=\linewidth]{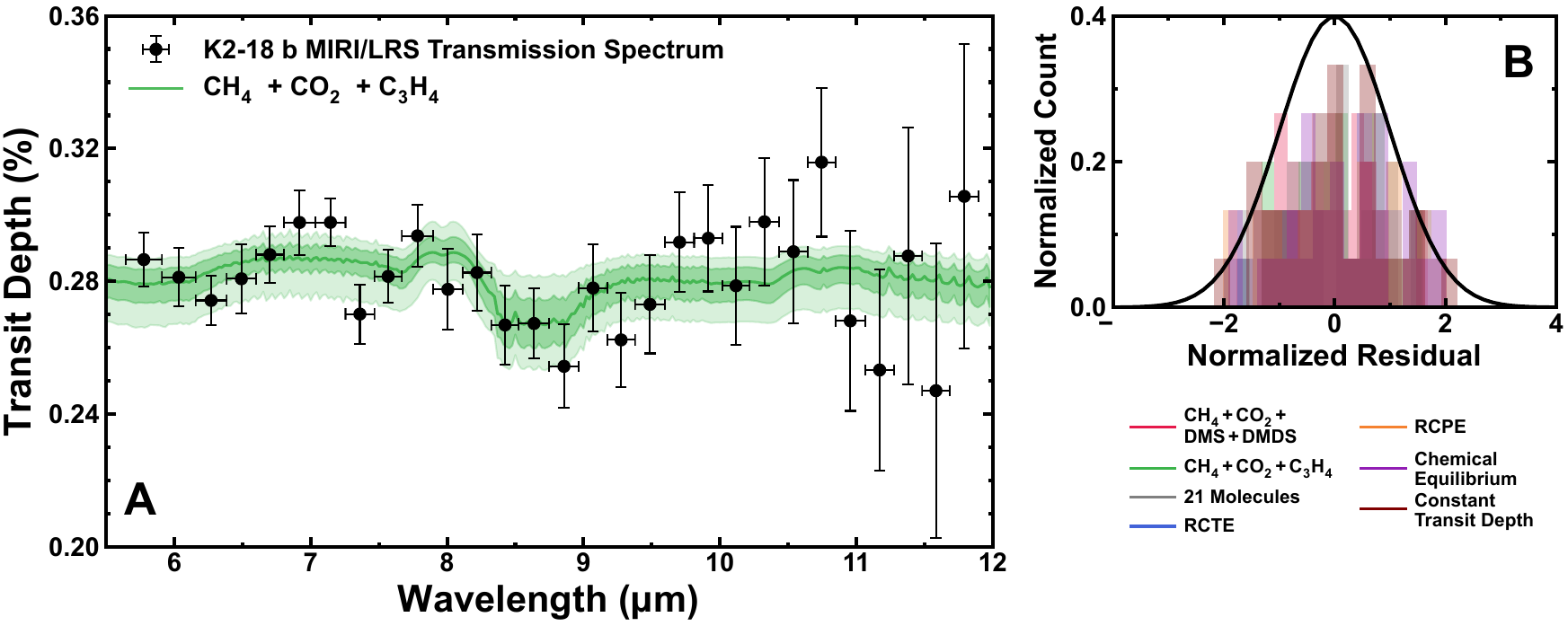}
     \caption{\textbf{JWST/MIRI transmission spectrum and model fits for K2-18 b.} \textbf{A} Transmission spectrum of K2-18 b reduced with the \texttt{JExoRES} pipeline\cite{Madhusudhan2025}. Black points show the mean transit depth per wavelength bin. Vertical error bars represent 1$\sigma$ standard deviations on the measured transit depth; horizontal bars indicate bin widths. The green line
     shows the median model including CH$_4$, CO$_2$ and C$_3$H$_4$ (propyne) with shaded bands denoting the 1$\sigma$ and 2$\sigma$ posterior intervals.
     \textbf{B} Normalized residuals ([data-model]/error) for parametric models including combinations of CH$_4$, CO$_2$, C$_3$H$_4$, DMS, and DMDS; self-consistent models considering radiative-convective-thermochemical equilibrium (RCTE) and radiative-convective-photochemical equilibrium (RCPE); a chemical equilibrium model where temperature structure is freely retrieved; and a constant transit depth (flat line) model. The black curve shows the expected normal distribution for comparison.} \label{fig:spectrum}
\end{figure*}

\paragraph{A Case Study: The MIRI/LRS Transmission Spectrum of K2-18~b}

We further explore the case of K2-18~b, now focusing on the JWST/MIRI transmission spectrum\cite{Madhusudhan2025}, to illustrate the risks of interpreting Bayesian model comparisons as gas detections. We find that the observations are consistent with a featureless spectrum, with any apparent signals being indistinguishable from statistical noise (see Methods, Extended Data Figure \ref{fig:ED1}). We also perform detailed atmospheric modeling as is routinely performed in the field, even for low signal-to-noise observations with weak spectral features\cite{Moran2023,BelloArufe2025,Schlawin2024,May2023}. We emphasize that consistency with the null hypothesis of a flat spectrum (see Extended Data Figure \ref{fig:ED2}) does not preclude atmospheric interpretation, but rather provides critical context for assessing the robustness of any inferences drawn from the data. Likewise, the preference of an atmospheric model over a featureless/constant transit depth model in Bayesian comparison alone does not reject the null hypothesis. 

We analyze the transmission spectrum with atmospheric models ranging in complexity from fully parametric approaches that independently fit the chemical abundances and pressure–temperature structure without enforcing physical self-consistency (commonly referred to as `free retrievals'), to those that couple radiative–convective and thermochemical equilibrium with disequilibrium processes such as photochemistry and vertical mixing (see Methods, Extended Data Figure \ref{fig:ED3}). Each approach has limitations: for example, equilibrium models may fail to capture atmospheres strongly influenced by disequilibrium chemistry, while fully parametric models may produce solutions that are not physically plausible\cite{Welbanks2024}.

We find that each modeling framework provides an adequate fit to the spectrum regardless of its underlying assumptions (see Methods, Extended Data Figure \ref{fig:ED4}). We first consider physically motivated models that assume (i) radiative–convective–thermochemical equilibrium (RCTE), (ii) radiative–convective–photochemical equilibrium (RCPE), and (iii) chemical equilibrium. The model residuals, shown in Figure~\ref{fig:spectrum}, are consistent with a normal distribution in all three cases. This indicates that the data do not reject the physics encapsulated within these assumptions. We conclude that the observations lack the constraining power necessary to distinguish between these physically distinct scenarios. Moreover, no model obtains meaningful constraints on the atmospheric properties, with posterior distributions closely reflecting the input priors. This emphasizes that an adequate model fit does not necessarily imply that the inferred properties from a given model reflect the true atmospheric state of the planet, particularly when multiple models remain compatible with the data.

We also consider a more flexible approach that allows the volume mixing ratios (i.e., abundances) of each chemical species to vary freely and independently. These models can attribute specific spectral features to particular chemical components. The Bayesian model comparison framework was originally introduced for this methodology\cite{Benneke2013}.  Practitioners must decide which chemical species to include based on prior expectations of their plausibility in the atmosphere in question. This choice is limited by both the number of species for which opacity data are available (several hundred\cite{Gordon2022}) and computational cost\cite{Fortney2021}. Here we include 21 chemical species motivated by theoretical calculations\cite{Moses2013a, Madhusudhan2019a}, observations of Solar System planet atmospheres\cite{Moses2000}, and previous studies of K2-18~b\cite{Madhusudhan2023} (see Methods, Extended Data Figure \ref{fig:ED3}). We include the proposed biosignature gases dimethyl sulfide (DMS) and dimethyl disulfide (DMDS), whose tentative presence has been reported in the atmosphere of K2-18~b\cite{Madhusudhan2025}.

\begin{figure*}[!htbp]
     \centering
     \includegraphics[width=\linewidth]{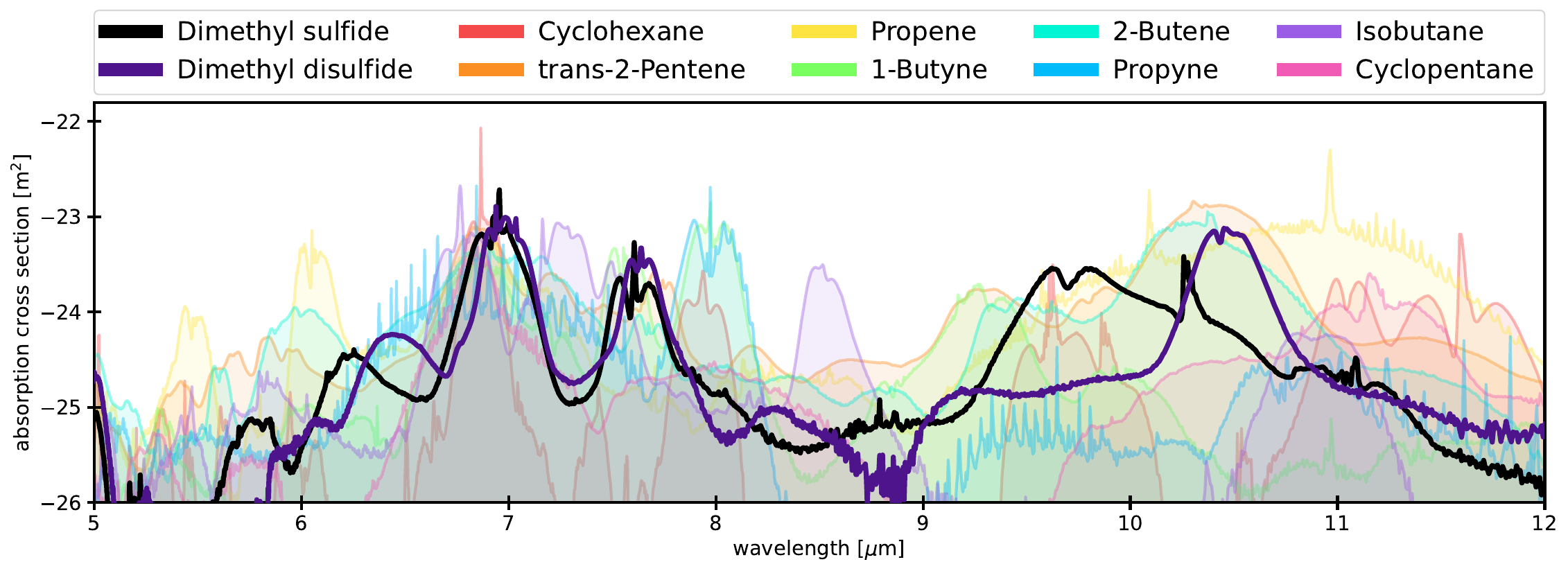}
     \caption{\textbf{Absorption cross-sections of dimethyl sulfide, dimethyl disulfide, and several hydrocarbons in the JWST/MIRI wavelength range.} Absorption cross-sections between 5–12$\mu$m for dimethyl sulfide (black), dimethyl disulfide (purple), and several hydrocarbons (colored lines) used in the model exploration. Prominent overlapping absorption features occur near $7\mu$m and $10$-$11\mu$m. where multiple species exhibit comparable opacity.} \label{fig:xsec}
\end{figure*}

These flexible retrievals of the K2-18\,b MIRI observations do not yield conclusive results, consistent with our findings with the physically motivated models. Multiple gas combinations provide adequate fits to the observed spectrum. To quantify the support for individual chemical species, we compare the Bayesian evidence of the full model (including 21 chemical species) against that of a set of nested models, each neglecting one species at a time\cite{Benneke2013}. Among the 21 species considered, only five led to a change in Bayesian evidence corresponding to a model preference $\geq$1.2$\sigma$ ($\mathcal{B} \leq 1.1 \pm 1.0$) for \texttt{JExoRES} and only two for \texttt{JexoPipe}, with all significances well below 2$\sigma$, including CH$_4$, CO$_2$, DMS, and DMDS (Extended Data Table~\ref{tab:big_retrieval}). The best-fit solutions from both the full model and the nested models yield $\chi^2/N_{\rm data} \leq 0.65$ for \texttt{JExoRES} and $\chi^2/N_{\rm data} \leq 0.72$ for \texttt{JexoPipe}, which are indicative of overfitting in all cases. In the full model, the only species whose posterior deviated noticeably from the prior for both the \texttt{JExoRES} and \texttt{Jexopipe} reductions was propyne (C$_3$H$_4$), a hydrocarbon identified in the atmospheres of Titan\cite{Maguire1981}, Saturn\cite{deGraauw1997}, Jupiter\cite{Fouchet2000}, Uranus\cite{Burgdorf2006}, and Neptune\cite{Meadows2008}.

We emphasize that this model implicitly assumes that no other gases beyond these 21 are present in the atmosphere of K2-18~b. Any inferences from this model must be interpreted within that context. As seen in Figure~\ref{fig:xsec}, several chemical species not considered in this 21-species model have cross-sections that overlap significantly in the wavelength range of these JWST/MIRI observations. Some of the prominent DMS and DMDS features, especially those at $\sim7\,\mu$m and $\sim 10.5\,\mu$m, are similar for  hydrocarbons not included in this 21 species model. Given these spectroscopic degeneracies and the low signal-to-noise ratio of these JWST/MIRI observations, multiple chemical species are likely to provide a reasonable fit and exacerbate the attribution problem. Indeed, we find that expanding this model to include 29 absorbers yields similar results (see Methods).

We now consider the reduction to a smaller model, as performed in the previous analysis of K2-18~b\cite{Madhusudhan2025}, to assess corresponding model preferences. In that work, an atmospheric model with 32 parameters (20 molecular abundances) was initially considered and then reduced to a smaller model with 16 parameters (4 molecular abundances) including only CH$_4$, CO$_2$, DMS and DMDS.  The presence of CH$_4$ and CO$_2$ was informed by prior observations\cite{Madhusudhan2023}. By comparing this four-gas model to a two-gas model excluding DMS and DMDS, the authors found a model preference between 2.9$\sigma$ and $3.4\sigma$ in favor of the inclusion of the sulfur species. We independently reconstruct this model configuration (see Methods, Extended Data Table \ref{tab:EDTable2}) and recover a model preference of $3.0\sigma$ for the inclusion of DMS or DMDS relative to the model containing only CO$_2$ and CH$_4$ --- a model comparison approach that deviates from standard practice of comparing nested models (see Methods, Extended Data Table \ref{tab:EDTable3}, Extended Data Figure \ref{fig:ED5}).

We extend this analysis with the aim of testing the robustness of the $\sim$3$\sigma$ confidence in the presence of DMS and DMDS. Specifically, we determine whether additional species that were not considered in the original analysis could equally or better explain the observations. We built 90 total models, each of which includes CH$_4$, CO$_2$, and a single hydrocarbon, covering all available HITRAN cross-sections\cite{Sharpe2004,Gordon2022}. We consider hydrocarbons for the following reasons: (i) CH$_4$ is the dominant carbon carrier at low temperatures ($\sim 250$~K)\cite{Moses2013a}, and several hydrocarbons are products of CH$_4$ photochemistry and thermochemistry at these temperatures (see Methods, Extended Data Figure \ref{fig:ED3}), (ii) several hydrocarbons have been identified in the H-rich atmospheres of solar system bodies including Titan and Neptune\cite{Waite2005,Meadows2008}, and (iii) many hydrocarbons have spectral features in the MIRI/LRS wavelength range\cite{Sousa-Silva2019} (Figure \ref{fig:xsec}).

We find that almost every hydrocarbon increases the Bayesian evidence relative to the CH$_4$+CO$_2$-only model, with 60 models showing a preference greater than 2$\sigma$ for at least one reduction of the data (Extended Data Table~\ref{tab:EDTable4} and Extended Data Table~\ref{tab:EDTable5}). The best-performing model, shown in Figure~\ref{fig:spectrum}, includes propyne (C$_3$H$_4$), yielding a $\gtrsim 3.1\sigma$ preference over the CH$_4$+CO$_2$-only model and a higher Bayesian evidence than the model including DMS and DMDS for both reductions. This indicates that, even within the limited framework of comparing a given species or pair of species against a model with CH$_4$+CO$_2$ only, DMS and DMDS are not the most preferred gases. By expanding the model space to include additional absorbers, we find that combining a small number of hydrocarbons with H$_2$O can lead to scenarios where the inclusion of DMS and DMDS cause the Bayesian evidence to decrease, meaning that model comparison favors the absence of these species (see Methods).

It is critical to note that the 3.1$\sigma$ preference for propyne is relative to a specific reference model, and does not constitute a detection of propyne (or any other species). This is analogous to the previous $\sim3.0\sigma$ preference for DMS and DMDS.  The apparently high detection significance of an individual gas is misleading in small model spaces where fewer species are available to explain the observations\cite{Benneke2013,Trotta2008a}. When more gases are included, degeneracies between overlapping absorption cross-sections preclude unique attribution of features (Figure \ref{fig:xsec}). Model preferences derived from limited comparisons must be interpreted with extreme caution, particularly when the overall constraining power of the data is weak.

\paragraph{Towards Reliable Atmospheric Characterization}\label{sec:conclusion}

Bayesian model comparison is a powerful framework for evaluating competing hypotheses, but does not by itself constitute robust detection of chemical species or atmospheric phenomena. Model preferences are always relative to a chosen reference model and are sensitive to the structure of the explored model space. Quoted significances are arbitrary, if not meaningless, without acknowledgment of their relative nature. We highlight that detection significances can be strongly affected by incomplete or constrained model assumptions, especially in the low-SNR regime. We do not suggest a specific threshold for detection, as any recommendation is meaningless without clear acknowledgment of its relative nature. However, as the degree of confidence in any claimed detection decreases --- either on a signal-to-noise ratio or model comparison basis --- the more important  context and the \textit{a priori} plausibility of the species should become.

As a case study of this phenomenon, we find that the JWST/MIRI transmission spectrum of K2-18~b  does not constrain the planet's atmospheric properties. We expanded the model space beyond the previously considered species, finding that other molecules, such as propyne (C$_3$H$_4$), provide comparable or greater Bayesian preference for the observed features than the proposed biosignatures DMS and DMDS\cite{Madhusudhan2025}. Propyne is photochemically plausible\cite{Moses2000}, has been detected in hydrogen-rich solar system planet atmospheres\cite{Burgdorf2006}, and has been suggested to be abundant on temperate sub-Neptunes, including K2-18~b\cite{Huang2024}. However, no molecules, including propyne, are identified with high confidence. We note that recent studies of the combined NIRISS + NIRSpec + MIRI spectrum of K2-18~b\cite{Luque2025,Hu2025} found evidence for CH$_4$ and CO$_2$ (c.f., ref\cite{Schmidt2025}) due to their larger wavelength coverage than MIRI only, but they did not find statistically significant indications of DMS or DMDS. These analyses did not consider C$_3$H$_4$, but given the signal-to-noise ratio of the present data, alongside differing interpretations of the presence and abundance of major species including CH$_4$, CO$_2$ and H$_2$O\cite{Schmidt2025,Jaziri2025}, we expect that unambiguous identification of a less prominent molecule such as C$_3$H$_4$ would require a larger observational investment, possibly requiring observations from future Extremely Large Telescopes\cite{Palle2025}. 

Interpreting exoplanet spectra remains one of the most challenging frontiers in planetary science. Atmospheric retrieval results should not be equated with chemical detections without context. A true understanding of atmospheric properties requires the synthesis of diverse modeling frameworks, from flexible parameterized to physically consistent, chemically coupled models. As the field matures, so too must our treatment of uncertainty. Future progress will require higher signal-to-noise observations, broader wavelength coverage to break spectral degeneracies, and continued development of rigorously benchmarked modeling tools. The challenges discussed here are not failures, but signs of a community moving towards deeper, more reliable inference. In the near term, for low-SNR spectra where unique chemical attribution is not possible, the most reliable approach is to report model comparisons only as measures of relative model adequacy, and to always contextualize them with additional tests; both to evaluate whether features are distinguishable from noise and to assess whether the proposed interpretation is physically and chemically plausible relative to competing hypotheses. Importantly, this is not a dispute between Bayesian and frequentist statistics; rather, we argue that robust atmospheric characterization depends on contextualizing results through multiple, complementary metrics, each addressing different questions. Just as multiple modeling approaches offer distinct insights and limitations, it is the overreliance on any single metric that risks obscuring what we actually know, and how confident we can be. In the end, a more systematic, physically motivated, and careful approach --- one that explicitly acknowledges the limitations of our models and the impact of what we choose to include or omit --- is essential to understanding these worlds.

\bibliographystyle{sn-standardnature}

\backmatter

\section*{Methods} \label{sec:methods}
\renewcommand{\figurename}{Extended Data Fig.}
\renewcommand{\tablename}{Extended Data Table}
\renewcommand{\theHfigure}{Extended Data Fig.~\arabic{figure}}
\renewcommand{\theHtable}{Extended Data Table \arabic{table}}
\setcounter{figure}{0}
\setcounter{table}{0}


\section*{Identification of Features in a Transmission Spectrum} \label{sec:flat_line_test}

As discussed in the main text, one of the key considerations when interpreting exoplanet atmospheric spectra is whether statistically significant features are present to interpret. Bayesian model comparisons aim to quantify the relative preference for one model over another given the data. A distinct but complementary analysis may be performed within a frequentist framework, which seeks to assess the probability that the effect of interest is not present in the data. In the context of atmospheric characterization, this null hypothesis typically corresponds to a featureless spectrum. This framework deviates significantly from the Bayesian perspective, where prior knowledge and expectations can be incorporated. For instance, within a Bayesian model comparison, we may ask: given that we know the planet has an atmosphere, which model --- an atmospheric model or a featureless model --- is more strongly supported by the data?

Ruling out the flat line hypothesis is not necessarily a prerequisite for interpreting atmospheric spectra. As the field continues to push towards characterizing smaller and cooler exoplanets, many observations will inevitably probe thin atmospheres with small scale heights, leading to weak or even statistically insignificant spectral features. Null hypothesis tests therefore provide important additional context to qualify any inferences drawn from the data. For instance, clarifying that the data quality is insufficient to strongly reject a flat spectrum can alert practitioners to proceed with caution and to avoid overinterpreting the strength of any Bayesian model comparison. Here, we demonstrate several frequentist approaches that may be used to assess whether a spectrum is statistically consistent with the null hypothesis of a featureless spectrum. We analyze mid-infrared transmission spectroscopy of K2-18\,b obtained with the JWST/MIRI Low-Resolution Spectrograph (LRS) as part of General Observer Program 2722 (PI: N.~Madhusudhan). The observations span the 5--12\,$\mu$m range and were conducted in slitless mode using the LRS double prism. We use two independently reduced transmission spectra from the same study\cite{Madhusudhan2025}. 

\subsection*{$\chi^2$ Flat Line rejection and Posterior Data Instances}\label{sec:chisq}

We perform a $\chi^2$ flat line rejection test\citeApp{Bevington} on both the \texttt{JExoRES} and \texttt{JexoPipe} reduced spectra against their mean transit depths of $2807$ ppm and $2817$ ppm, respectively. This is a hypothesis test, where the null hypothesis is that the observed spectrum arises from a featureless atmosphere with only random noise. We compute $\chi^2_{\nu}$, or $\chi^2$ divided by the degrees of freedom $\nu=N_{\rm data}-N_{\rm params}$. In this case, $N_{\rm params}=1$ (the mean transit depth). For \texttt{JExoRES}, we measure $\chi^2=30.0$, or $\chi^2_{\nu}=1.07$. and for \texttt{JexoPipe}, we measure $\chi^2=34.1$, or $\chi^2_{\nu}=1.22$. Based on this $\chi^2$ analysis, the \texttt{JExoRES} and \texttt{JexoPipe} spectra reject a flat line spectrum at low confidence, with $p$-values of 0.36 and 0.20 (or $0.9\sigma$ and $1.3\sigma$), respectively. This result has been reported previously \citeApp{Taylor2025}.

The null hypothesis of a featureless spectrum is also not rejected when considering the maximum-likelihood solution for a model in which the transit depth ($\Delta$) is treated as a free parameter. Fitting this model, we find $\Delta=2819$ ppm with $\chi^2 = 29.7$, or $\chi^2_{\nu}=1.06$ ($p$-value = 0.38) and $\Delta=2848$~ppm with $\chi^2 = 32.4$, or $\chi^2_{\nu}=1.16$ ($p$-value = 0.26) for the \texttt{JExoRES} and \texttt{JexoPipe} reductions, respectively. In both cases, the results lie well within the expected $1\sigma$ range for $\chi^2_\nu$ in our dataset, approximately $0.74$ to $1.26$\citeApp{Andrae2010}. 

\begin{figure*}
    \centering
    \includegraphics[width=\linewidth]{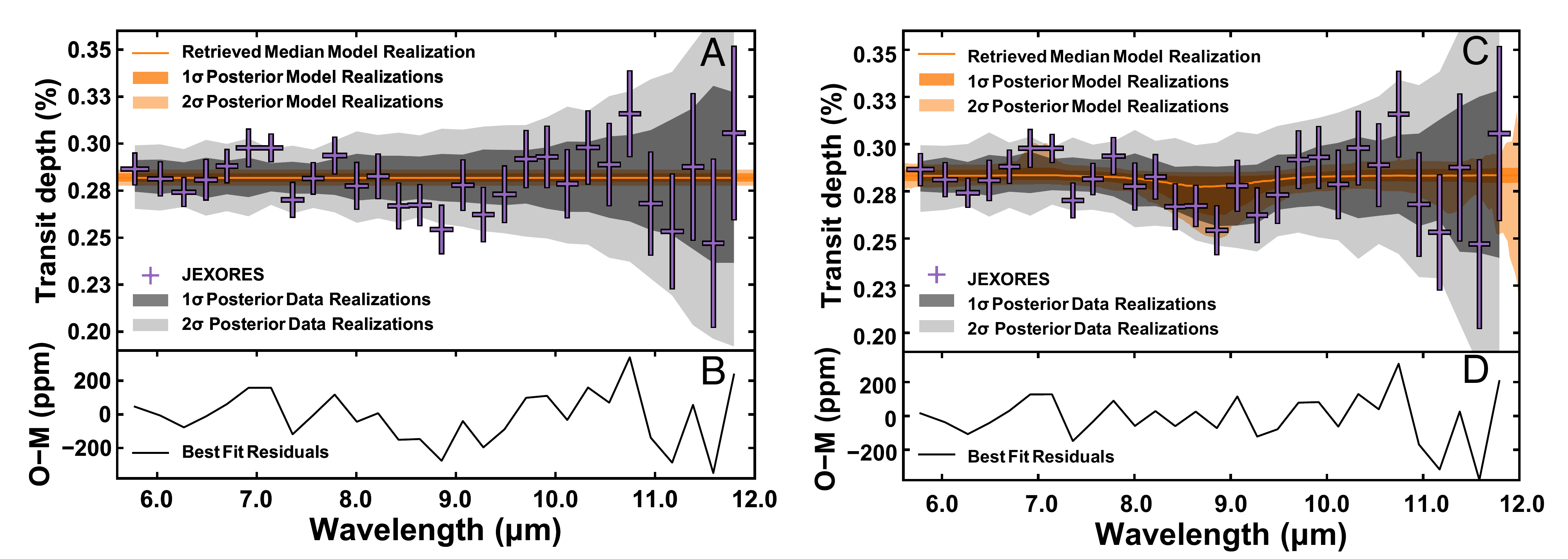}
    \caption{\textbf{Posterior model and data realizations for MIRI observations of K2-18 b.} \textbf{A)} Posterior data (gray) and model (orange) realizations generated from a flat-line model. Purple points indicate the median retrieved transit depth in each wavelength bin; vertical error bars represent the uncertainty in transit depth and horizontal bars indicate the bin width. \textbf{B)}  Residuals between the data realizations and the best-fit flat-line model. \textbf{C)} As panel A, but for the model including a single-kernel Gaussian Process (GP) noise component. The GP identifies correlated structure near 7 µm and 8.5–9 µm, but this additional complexity is not statistically preferred over the flat-line model with white noise.    \textbf{D)} Residuals between the data realizations and the best-fit GP model.}  
    \label{fig:ED1}
\end{figure*}

We further obtain posterior samples of the transit depth under the flat spectrum model, assuming a uniform prior between 0 and 5686~ppm and using nested sampling\citeApp{Buchner2014}. We recover posterior constraints on the transit depth of $2820_{-21}^{+22}$~ppm and $2847_{-22}^{+24}$~ppm for the \texttt{JExoRES} and \texttt{JexoPipe} reductions, respectively. Extended Data Figure~\ref{fig:ED1} shows posterior realizations from the flat spectrum model overplotted on the \texttt{JExoRES} observations. The consistency between the posterior data realizations and the JWST observations provides intuition for why the featureless null hypothesis cannot be statistically rejected, and highlights that any apparent deviations from flatness are indistinguishable from statistical noise at the current precision.

\begin{figure}
    \centering
    \includegraphics[width=1.\linewidth]{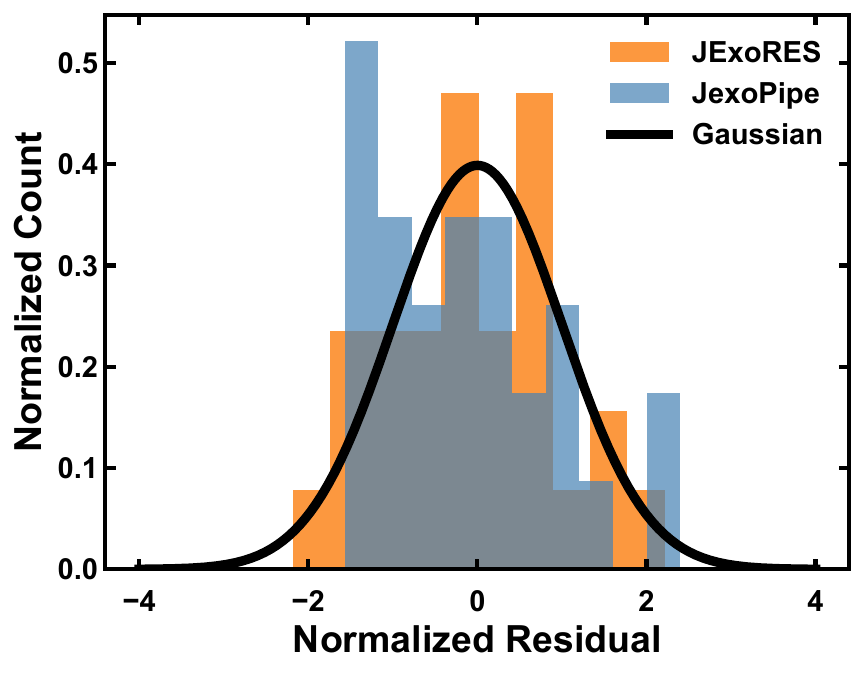} \caption{\textbf{Normalized residuals of the \texttt{JExoRES} and \texttt{JexoPipe} spectra relative to flat-line models.} Histograms of normalized residuals, defined as $([{\rm data} - {\rm model}]/\sigma)$, for the \texttt{JExoRES} and \texttt{JexoPipe} reductions compared with their best-fit featureless spectra. Data represent single measurements per wavelength bin (n=1); bars show the frequency distribution of normalized residuals. The black curve indicates the standard Normal (Gaussian) distribution used for comparison.}
    \label{fig:ED2}
\end{figure}

\subsection*{Assessing Structure in Model Residuals}\label{sec:residuals-flat-line}

We perform the Kolmogorov-Smirnov (K-S) and the Anderson-Darling (A-D) tests on the data-model residuals around the constant transit depth maximum likelihood solution\citeApp{DAgostino1986}. For the A-D test, we report the adjusted A-D statistic to account for the small number of data points in the spectrum\citeApp{DAgostino1986}. Both tests compare normalized residuals between the observations and a model to a Normal distribution with mean $\mu=0$ and variance $\sigma^2=1$. Residuals consistent with a Normal distribution suggest that the variation about the model -- here, a featureless spectrum -- is likely due to random measurement error. The histograms of the normalized residuals for the \texttt{JExoRES} and \texttt{JexoPipe} data, against their corresponding maximum-likelihood flat spectra are shown in Extended Data Figure~\ref{fig:ED2}. For \texttt{JExoRES}, we find a K-S statistic of 0.07 with a $p$-value of 0.99, and an A-D statistic of 0.15 with a $p$-value of 0.96. For \texttt{JexoPipe} we find a K-S statistic of 0.13 with a $p$-value of 0.63, and A-D statistic of 0.39 with a $p$-value of 0.39. These high p-values indicate that the residuals are statistically consistent with being drawn from a standard Normal distribution. Therefore, both the K-S and A-D tests suggest that the featureless spectrum model is a statistically adequate explanation (i.e., not rejected) for both reductions.

\subsection*{Gaussian Process Modeling}\label{sec:gp-flat-line}

We further assess the possibility of statistically significant spectral features by repeating the flat line analysis using a Gaussian Process (GP) noise model\citeApp{Rasmussen2006}. Using a GP increases the flexibility of the model, allowing it to fit or marginalize over correlated features that would otherwise bias a simpler white noise model. Following the framework for exoplanet spectral analysis outlined by ref.\citeApp{Rotman2025}, we include a GP with a single local kernel, designed to capture localized deviations from a flat spectrum, such as unresolved absorption features. This approach enables an assessment of whether any apparent deviations from flatness are statistically significant given the noise properties of the data. Our approach complements the Gaussian feature analysis in ref.\citeApp{Taylor2025}, which fits parametric Gaussian functions at specific wavelengths to test for localized absorption features. In contrast, our Gaussian Process framework models the entire spectrum non-parametrically, enabling the identification of broad or structured deviations without assuming a specific functional form. Our results are shown in Extended Data Figure \ref{fig:ED1}.

The GP analysis identifies two regions of the spectrum where the addition of a local kernel may improve the model fit: one at $\sim8.5$--$9~\mu$m, where the observations fall below the median flat-line model, and another near $\sim7~\mu$m, where the observations lie above it. A model comparison between this more flexible GP model and the featureless model does not yield a statistically significant preference. Overall, the retrieved transit depths from both models are consistent with each other within the 68.26-percentile intervals of the posterior distributions. These results suggest that, while the data are broadly consistent with a featureless spectrum, chemical absorbers with spectral features near $\sim7~\mu$m and $\sim8.5$--$9~\mu$m may provide an improved fit.

\section*{Atmospheric Modeling}

We interpret the JWST/MIRI spectrum of K2-18~b using a series of atmospheric models, and perform parameter estimation using models with varying degrees of complexity and physical plausibility. Our model frameworks are varied in terms of their assumptions regarding the chemical composition and pressure-temperature profile of the planet. Our self-consistent models impose the most stringent assumptions on the atmosphere, since radiative-convective equilibrium is used to compute the pressure-temperature profile given the incident stellar flux, internal heat flux, and opacity structure, meaning that the pressure-temperature profile and chemistry are coupled. Chemical abundances are computed from elemental abundances assuming thermochemical or photochemical equilibrium. Our chemical equilibrium models retain this computation of elemental abundances from thermochemical equilibrium, but we decouple the pressure-temperature profile, allowing it to be determined independently. In the least restricted framework, we make no assumptions about the chemical composition or pressure-temperature profile, allowing each chemical species to be an independent free parameter along with free parameters for the pressure-temperature profile.

Any approach will have certain limitations: for example, the assumption of chemical equilibrium may fail to explain an atmosphere where strong disequilibrium effects, such as photochemistry or vertical mixing, are present. It may also fail to account for as-yet unknown physical, chemical or even biological processes. On the other hand, a model where all parameters are independent and can vary freely could result in unphysical or highly implausible models being used to explain the observations. Besides fitting the observations, we further model the atmosphere of K2-18~b using two independent chemical kinetics models to investigate the plausible chemical species in the atmosphere of this mini-Neptune given a metallicity and carbon-to-oxygen ratio. 

In all models described below, we fix the planet mass to 8.63$M_{\oplus}$\citeApp{Cloutier2019a} and radius to 2.61$R_{\oplus}$\cite{Benneke2019}, and the stellar radius to 0.4445$R_{\odot}$\cite{Benneke2019}. We assume a one-dimensional atmosphere in which parameters can vary with pressure (height), but not latitude or longitude. Stellar heterogeneity has not been found to impact transmission spectra of K2-18~b\cite{Benneke2019,Schmidt2025} and we do not consider it in this work, although we note that in general, stellar heterogeneity can imprint signals similar to molecular features, adding an additional potential degeneracy to be considered when analysing transmission spectra\citeApp{Rackham2018}.

\subsection*{Self-Consistent Models}\label{sec:Grids}

\texttt{ScCHIMERA} is a 1D radiative-convective equilibrium framework for exoplanet atmospheres\citeApp{Bell2023c,Piskorz2018,Mansfield2021,Iyer2023} and we closely follow the implementation in ref.\cite{Welbanks2024}. \texttt{ScCHIMERA} iteratively solves for the radiative-convective equilibrium pressure-temperature profile based on the two-stream source function method\citeApp{Toon1989} and a Newton-Raphson iteration scheme\citeApp{McKay1989}. Models are computed in a log-uniform pressure grid ranging from $10^{-6}$ bar to $10^{3}$ bar in steps of 0.1 dex. Thermochemical equilibrium abundances are computed in each pressure layer using the \texttt{NASA CEA2} routine, which minimizes the Gibbs free energy\citeApp{Gordon1994}. We also account for elemental rainout due to condensate formation. Abundances are computed for numerous molecular species, but we include opacity sources for the following: H$_2$-He collision-induced absorption\citeApp{Karman2019}, H$_2$O\citeApp{Polyansky2018}, CO\citeApp{Li2015}, CO$_2$\citeApp{Huang2012}, CH$_4$\citeApp{Hargreaves2020}, NH$_3$\citeApp{Coles2019}, H$_2$S\citeApp{Azzam2016}, HCN\citeApp{Harris2006a}, C$_2$H$_2$\citeApp{Chubb2020}, Na\citeApp{Allard2019}, K\citeApp{Allard2016}, and SO$_2$\citeApp{Underwood2016a}. Model spectra are generated from 5-12~$\mu$m at $R=100\,000$. We adopt a fixed incident stellar flux (derived from a \texttt{PHOENIX} synthetic spectrum\citeApp{Husser2013} with $T_{\text{eff}} = 3464\ \text{K}$ and $\log_{10} g = 4.786\ \log_{10}(\text{cm s}^{-2})$\citeApp{Stassun2019}). The calculation of disequilibrium processes is done by coupling the radiative-convective equilibrium solver described above with the kinetics code \texttt{VULCAN}\citeApp{Tsai2017,Tsai2022}, as previously described in refs.\citeApp{Bell2023c}$^,$\citeApp{Welbanks2024}, with a single value for eddy mixing ($\log_{10}K_{zz}=7$) and intrinsic temperature (50~K based on evolutionary models\citeApp{Lopez2014}).


\begin{figure*}
    \centering
    \includegraphics[width=\linewidth]{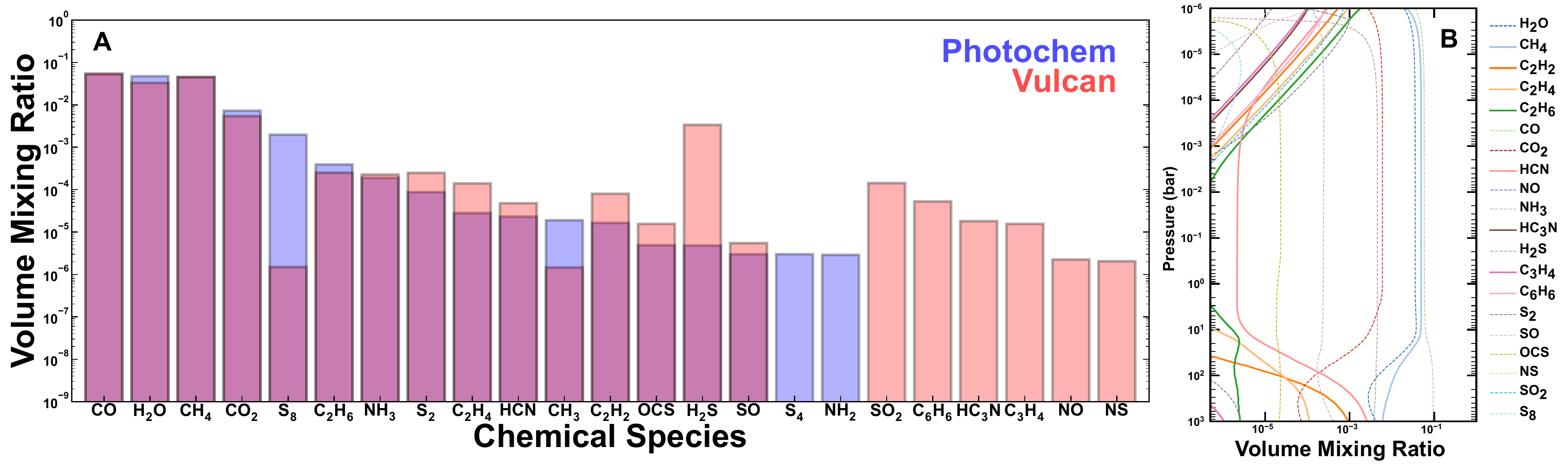}
    \caption{\textbf{Predicted chemical composition of a K2-18 b-like atmosphere from radiative–convective–photochemical and kinetics models.} \textbf{A)} Average volume mixing ratios of gases with abundances exceeding 1 ppm in the 1~mbar–1~$\mu$bar region, computed using two independent photochemical kinetics models, Photochem (blue) and Vulcan (red) for an atmosphere with metallicity [M/H] = +2.25 and C/O = 0.8. Only species with infrared-active bands are shown. \textbf{B)} Vertical abundance profiles from the 1D radiative–convective–photochemical equilibrium (1D-RCPE) grid for the same composition ([M/H] = 2.25, C/O = 0.8) assuming full heat redistribution. Several hydrocarbons and sulfur-bearing products are present at non-negligible abundances in the observable atmosphere. These species arise from coupled thermochemical, photochemical, and vertical-mixing processes within the modeled atmosphere.}
    \label{fig:ED3}
\end{figure*}

We generate two grids of models, one under radiative-convective-thermochemical equilibrium (RCTE), and another under radiative-convective-photochemical equilibrium (RCPE). All models vary across metallicity ($1.00 \leq \text{[M/H]} \leq 2.75$ in steps of 0.25, in which M denotes all non-H/He elements and [ ] denotes log$_{10}$ relative to solar), the carbon-to-oxygen ratio ($0.1 \leq \text{C/O} \leq 0.7$ in steps of 0.1 and $0.70 < \text{C/O} \leq 1.00$ in steps of 0.05), and day-night heat redistribution factors ($f = 0.452,\ 0.686,\ 1.000,\ 1.411,\ 1.938$). We produce 520 RCTE and 520 RCPE models in total. Elemental abundances available in the atmosphere are determined via [M/H] and C/O. We use \texttt{PyMultiNest}\citeApp{Buchner2014}, a Python implementation of \texttt{MultiNest}, a variant of the Nested Sampling algorithm\citeApp{Skilling2004a,Feroz2009a}, with 2000 live points for parameter estimation. During the Bayesian inference step, we additionally add free parameters for the reference radius $R_p$ (uniform prior, 0.5$\times$ to 1.5$\times$ the adopted radius of K2-18~b\cite{Benneke2019}), the reference pressure $P_{\rm ref}$ (log-uniform prior between $10^{-8.0}$~bar and $10^{1.3}$~bar), and inhomogeneous clouds and hazes following the one-sector parameterization of ref.\citeApp{Welbanks2021} as described below for the parametric models and in previous works\citeApp{Bell2023c}$^,$\cite{Welbanks2024}, with the distinction that we parameterize the effect of an optically thick cloud deck with a cloud opacity $\kappa_\text{cloud}$ (log-uniform prior between  $10^{-50}$ and  $10^{-20}$).

The retrievals using both 1D-RCTE and 1D-RCPE yield good fits to the data, as shown in Figure~\ref{fig:spectrum} and Extended Data Figure~\ref{fig:ED4}, where the median model produces residuals that are statistically consistent with a normal distribution\citeApp{Andrae2010}. However, these retrievals do not yield meaningful constraints, as the retrieved posterior distributions closely reflect the input priors. The one exception is the planetary radius at the reference pressure, which is retrieved to $R_p = 0.96 \pm 0.02 \times$ K2-18~b's input radius. This, however, results from the $R_p$--$P_{\rm ref}$ degeneracy\citeApp{Welbanks2019a} and reflects the choice of pressure grid and $P_{\rm ref}$ prior. Extended Data Figure \ref{fig:ED3} shows the abundance profiles for a composition of [M/H] $=2.25$, C/O $=0.8$, and full redistribution from the 1D-RCPE grid, one of the multiple solutions permissible by the data. 

To probe the relative importance of each absorber, we perform additional retrievals using the same model grids but with the abundance of individual gases fixed to $10^{-50}$ during sampling, as in previous work\cite{Welbanks2024}. This emulates a nested model comparison framework within the grid-fitting approach. For each species, we compare the resulting model to the full reference model that includes all absorbers. No significant model preference is found for any species. In summary, these self-consistent models are statistically sufficient to explain the data. 

\subsection*{Chemical Equilibrium Models}\label{sec:Picaso}

We use the modeling framework \texttt{PICASO}\citeApp{Mukherjee2023} to build models assuming equilibrium chemistry, but with a decoupled pressure-temperature profile. We follow the model setup of the self-consistent model fitting described above in this framework as well unless explicitly specified here. Following the same method as ref.\cite{Madhusudhan2025}, we use the $T(P)$ profile parameterization from ref.\citeApp{Madhusudhan2009a} to model the temperature structure of the planet. We divide the atmosphere in 150 plane-parallel layers with pressures that are logarithmically spaced between 10$^{-9}$-1 bars. This parametrization has six input parameters: $T_0$, $\alpha_1$, $\alpha_2$, $P_1$, $P_2$, and $P_3$. We use the same priors as ref.\cite{Madhusudhan2025} on these parameters. We simulate the chemical composition of the atmosphere based on the $T(P)$ profile assuming thermochemical equilibrium using two input parameters: metallicity ([M/H]) and C/O. The [M/H] parameter is allowed to vary between $-1$ and $+3.0$ whereas the C/O is allowed to vary between 0.0458 and 0.916. We include the effects of aerosols in our retrieval model using three parameters: cloud opacity $\kappa_{\rm cld0}$, a size parameter ($a$), and a slope parameter ($s$). The aerosol optical depth in a vertical atmospheric column is gray and is given by $\tau=\kappa_{\rm cld0}\delta{P}/g$ for longer wavelengths that follow $\lambda\ge2\pi{a}$. For shorter wavelengths that follow $\lambda < 2\pi{a}$, the vertical aerosol optical depth is also given by $\tau=\kappa_{\rm cld}\delta{P}/g$, where  $\kappa_{\rm cld}= \kappa_{\rm cld0}({2{\pi}a}/{\lambda})^s$. Such an aerosol implementation aims to capture the gray scattering nature of aerosols for longer wavelengths and the aerosol-induced scattering slope at shorter wavelengths simultaneously. We allow $\log_{10}(\kappa_{\rm cld0})$ to vary between 0 and $-30$ in cgs units. The size parameter ($a$) is allowed to vary between 0.1-15~$\mu$m and $s$ is varied between 0 to 15. Following the treatment of ref.\cite{Madhusudhan2025}, we also allow for patchy clouds using the cloud fraction parameter $\phi$, which is allowed to vary between 0-1. The total spectrum is obtained by a linear combination of a clear and a cloudy atmosphere using (1-$\phi$) and $\phi$ as weights. We also retrieve a reference radius at a mbar pressure as well as an offset parameter between the data and the model to maintain similarity with ref.\cite{Madhusudhan2025}. The offset parameter is allowed to vary within $\pm100$ ppm. The reference radius is varied through the $x_r$ parameter, where the radius at 1~mbar is defined as $R_{\rm mbar}= R_p(1+x_r)$. We allow $x_r$ to vary between $\pm$0.1. All priors are uniform.

We compute the transmission spectrum between 3.85-14.1 $\mu$m using resampled opacities at a spectral resolution of R$\sim$15,000. We include gaseous opacities for C$_2$H$_2$, C$_2$H$_4$, C$_2$H$_6$, CH$_4$, CO, CO$_2$, H$_2$, H$_2$O, H$_2$S, HCN, N$_2$, N$_2$O, NH$_3$, O$_2$, O$_3$, OCS, and SO$_2$ within our model. The sources of opacities used in our framework are same as described in ref.\citeApp{Mukherjee2024}. We use the \texttt{PyMultiNest}\citeApp{Buchner2014} nested sampling tool to constrain the parameters of this model with 2200 live points and the default convergence criteria of \texttt{PyMultiNest}.

Overall, most retrieved parameters reflect their priors, indicating that the data do not provide strong constraints under the assumptions of this model. The retrieved metallicity and C/O remain consistent with the applied priors within 2$\sigma$ confidence intervals. The best-fit model with the highest likelihood has parameter values that are consistent within the 1$\sigma$ constraints on all the parameters except $T_0$. The value of $T_0$ in the best-fit model is 441 K. We also calculate the Bayes factor that can be attributed to the presence of a certain gas in the chemical equilibrium model by removing one gas at a time and rerunning a new ``removed gas'' fit to the data. We find that the logarithm of the Bayes factor (ln$(\mathcal{B})$) changes the most when H$_2$O is removed from the model setup with ln$(\mathcal{B})$= 1.5 in favor of the presence of H$_2$O absorption (equivalent to a 2.3$\sigma$ model preference in favor of H$_2$O absorption). No other model comparisons were significant.

\subsection*{Photochemical Products}\label{sec:photochemistry}

We use two independently developed 1D chemical kinetics models, \texttt{photochem} \& VULCAN\citeApp{Wogan2024,Tsai2017}, to find the most abundant gaseous molecules in the photosphere of a planet like K2-18~b that can be produced through three atmospheric processes: thermochemistry, atmospheric vertical mixing, and photochemistry. We use the $T(P)$ profile of the best-fitting self-consistent \texttt{ScCHIMERA} RCPE grid model as an input and assume a strength of vertical mixing $\log_{10}(K_{\rm zz}/{\rm cm}^2{\rm s}^{-1})=7$. Extended Data Figure \ref{fig:ED3} shows the average abundance of gases between 1 mbar and 1 $\mu$bar in the \texttt{photochem} and VULCAN models. We have assumed a metallicity of [M/H] $=+2.25$ and C/O $=0.8$ for these models and only show molecules that have average abundances higher than 1 ppm within the 0.001-1 mbar pressure range. We have not shown gases that are known to lack prominent infrared absorption features such as H$_2$, He, N$_2$, and other monoatomic species. A key finding from these models is the relatively high predicted abundances of hydrocarbons like C$_2$H$_6$, C$_2$H$_4$, CH$_3$, C$_2$H$_2$, C$_6$H$_6$, HC$_3$N, and C$_3$H$_4$ near the observable photosphere of the planet. These gases are produced through a combination of photochemistry and vertical dynamics in the atmosphere within these atmospheric models. A large inventory of S- bearing gases such as S$_2$, S$_8$, OCS, H$_2$S, SO$_2$, SO, and NS are also predicted in these models. These results motivate the species considered in our parametric models and the consideration of hydrocarbons in the atmosphere of K2-18~b.

\begin{table*}[ht]
\centering
\caption{\justifying \textbf{Bayesian model comparison for molecular species detection using \texttt{JExoRES} and \texttt{JexoPipe} reductions for a model with 21 gases.} Bayesian evidences ($\ln \mathcal{Z}$) and Bayes factors ($\ln \mathcal{B}$) are shown relative to the full model including all gases. Results are provided for the \texttt{JExoRES} (\textdagger) and \texttt{JexoPipe} (\textdaggerdbl) reductions. Reported uncertainties are $\pm0.05$ for $\ln \mathcal{Z}$ and $\pm0.07$ for $\ln \mathcal{B}$.}
\resizebox{\textwidth}{!}{%
\begin{tabular}{l|cc|cc|cc|cc|cc}
\hline
\textbf{Model} 
& \multicolumn{2}{c|}{$\boldsymbol{\ln(\mathcal{Z})}$} 
& \multicolumn{2}{c|}{$\boldsymbol{\ln(\mathcal{B})}$} 
& \multicolumn{2}{c|}{\textbf{Preference} ($\boldsymbol{\sigma}$)} 
& \multicolumn{2}{c|}{$\boldsymbol{\ln(\mathcal{L}_{\max})}$} 
& \multicolumn{2}{c}{$\boldsymbol{\chi^2 / N_{\mathrm{data}}}$} \\
& \textdagger  & \textdaggerdbl & \textdagger  & \textdaggerdbl & \textdagger & \textdaggerdbl & \textdagger & \textdaggerdbl & \textdagger & \textdaggerdbl \\
\hline
Full model & 213.07 & 208.01 & Ref & Ref   & Ref & Ref  & 221.46 & 216.66 & 0.59 & 0.62  \\
Remove SO$_2$ & 213.16 & 208.11 & -0.09 & -0.10  & -- & --  & 220.95 & 217.33 & 0.62 & 0.57 \\
Remove CS$_2$ & 213.14 & 208.06  & -0.07 & -0.05  & -- & -- & 221.63 & 217.05 & 0.58 & 0.59 \\
Remove C$_2$H$_4$ & 213.11 & 207.98 & -0.04 & 0.03 & -- & -- & 221.77 & 216.81 & 0.57 & 0.61 \\
Remove C$_2$H$_2$ & 213.11 & 208.10 & -0.04 & -0.09  & -- & -- & 221.11 & 216.66 & 0.61 & 0.62 \\
Remove NH$_3$ & 213.10 &  208.07 & -0.03 & -0.06 & -- & -- & 221.10 & 216.98 & 0.61 & 0.59 \\
Remove C$_3$H$_8$ & 213.10 & 207.98 & -0.03 & 0.03 & -- & -- & 221.75 & 216.95 & 0.57 & 0.60 \\
Remove CH$_4$S & 213.10 & 208.09  & -0.03 & -0.08 & -- & -- & 221.40 & 217.29 & 0.59 & 0.58 \\
Remove OCS & 213.09 & 208.10  & -0.02 & -0.09 & -- & -- & 221.66 & 216.61 & 0.58 & 0.62 \\
Remove CO$_2$ & 213.08 & 208.10 & -0.01 & -0.09 & -- & -- & 221.83 & 216.62 & 0.56 & 0.62  \\
Remove H$_2$S & 213.08 & 208.01 & -0.01 & 0.00 & -- & -- & 221.31 & 216.75 & 0.60 & 0.61 \\
Remove HCN & 213.05 & 208.01 & 0.02 & 0.00 & -- & -- & 221.44 & 216.98 & 0.59 & 0.59  \\
Remove C$_4$H$_{10}$ & 213.05 & 208.07 & 0.02 & -0.06 & -- & -- & 221.38 & 217.06 & 0.59 & 0.59 \\
Remove CO & 213.04 & 208.03 & 0.03 & -0.02 & -- & -- & 221.79 & 216.88  & 0.57 & 0.60 \\
Remove PH$_3$ & 213.04 & 208.03 & 0.03 & -0.02 & -- & --  & 222.04 & 216.81  & 0.55 & 0.61 \\
Remove H$_2$O & 213.02 & 208.06 & 0.05 & -0.05 & -- & --  & 220.86 & 216.70 & 0.63 & 0.61 \\
Remove C$_2$H$_6$ & 213.00 & 208.08 & 0.07 & -0.07 & -- & --  & 221.67 & 216.94 & 0.57 & 0.60 \\
Remove CH$_4$ & 212.98 & 208.00 & 0.09 & 0.01 & 1.2 & -- & 221.22 & 216.79 & 0.61 & 0.61 \\
Remove DMS & 212.97 & 207.84  & 0.10 & 0.17  & 1.2 & 1.3 & 221.14 & 215.92 & 0.61 & 0.67 \\
Remove C$_4$H$_8$ & 212.95 & 208.12  & 0.11 & -0.11  & 1.2 & -- & 221.64 & 217.22 & 0.58 & 0.58 \\
Remove DMDS & 212.91 & 208.01 & 0.16 & 0.00 & 1.3 & -- & 220.86 & 216.99 & 0.63 & 0.59 \\
Remove DMS, DMDS & 212.87 & 207.64 & 0.20 & 0.37 & 1.4 & 1.6 & 220.60 & 215.20 & 0.65 & 0.72 \\
Remove C$_3$H$_4$ & 212.74 & 207.67 & 0.33 & 0.34 & 1.5 & 1.5 & 221.79 & 216.86 & 0.57 & 0.60 \\
Remove C$_4$H$_{10}$, C$_3$H$_4$, C$_4$H$_8$ & 212.43 & 207.64  & 0.64 & 0.37 & 1.8 & 1.6  & 221.93 & 216.45 & 0.56 & 0.63 \\
\hline
\end{tabular}%
}
\label{tab:big_retrieval}
\end{table*}

\subsection*{Parametric Models}\label{sec:Free-Retrievals}

We explore models in which all chemical abundances are allowed to vary freely using the atmospheric modeling code \textit{Aurora}\citeApp{Welbanks2021}$^,$\cite{Nixon2024a}. As with models discussed above, we use \texttt{PyMultiNest}\citeApp{Buchner2014} for evidence computation and parameter estimation. The pressure-temperature profile consists of 6 free parameters\citeApp{Madhusudhan2009a}: $T_0$ (uniform prior, 100 to 500~K), $\alpha_1$, $\alpha_2$ (uniform priors, 0.02 to 2 K$^{-1/2}$), $P_1, P_2$ (log-uniform priors, $10^{-7}$ to $10$ bar) and $P_3$ (log-uniform prior, $10^{-2}$ to $10^2$ bar). The white-light radius $R_p$ (uniform prior, $0.5\times$ to $1.5\times$ the adopted radius of K2-18~b) and the reference pressure $P_{\rm ref}$ (log-uniform prior, $10^{-7}$ to $100$ bar) at this radius are also taken as free parameters. Our pressure grid consists of 100 uniform layers in log space between 10$^{-7}$ and 100 bar. We assume a H$_2$/He dominated atmosphere with a solar composition of $X_{\textnormal{He}} / X_{\textnormal{H}_2} = 0.17$\citeApp{Asplund2009a}. Our model includes opacity due to collision-induced absorption from H$_2$-H$_2$ and H$_2$-He interactions\citeApp{Richard2012}, as well as extinction from a range of chemical species (see \ref{tab:EDTable3}). Volume mixing ratios of these chemical species follow a log-uniform prior ranging from 10$^{-20}$ to $10^{-0.3}$, under the constraint that the sum of the mixing ratios of all non-H$_2$/He species must not exceed 1. We also include opacity due to clouds/hazes following a patchy cloud retrieval methodology\citeApp{Etangs2008a,Line2016a}. This prescription incorporates four additional free parameters into the model: the Rayleigh-enhancement factor $a$ (log-uniform prior, 10$^{-4}$ to 10$^{10}$), haze slope $\gamma$ (uniform prior, $-20$ to 2), cloud top pressure $P_{\rm cloud}$ (log-uniform prior, $10^{-7}$ to $100$ bar) and the fractional cloud coverage $\bar{\phi}$ (uniform prior, 0 to 1).

We constructed a model including opacity contributions from 21 molecules: H$_2$O\citeApp{Rothman2010a}, CH$_4$\citeApp{Yurchenko2014a}, NH$_3$\citeApp{Yurchenko2011a}, HCN\citeApp{Barber2014a}, CO\citeApp{Rothman2010a}, CO$_2$\citeApp{Rothman2010a}, SO$_2$\citeApp{Underwood2016a}, H$_2$S\citeApp{Azzam2016}, CS$_2$\cite{Sharpe2004,Gordon2022}$^,$\citeApp{Karlovets2021}, OCS\citeApp{Wilzewski2016}, PH$_3$\citeApp{Sousa-Silva2015}, C$_2$H$_6$S (DMS)\cite{Sharpe2004}, C$_2$H$_6$S$_2$ (DMDS)\cite{Sharpe2004}, C$_4$HS\cite{Sharpe2004}, C$_2$H$_2$\citeApp{Rothman2013}, C$_2$H$_4$\citeApp{Mant2018}, C$_2$H$_6$\citeApp{Rothman2013}, C$_3$H$_4$ (propyne)\cite{Sharpe2004}, C$_3$H$_8$ (propane)\citeApp{Sung2013}, C$_4$H$_8$ (2-butene)\cite{Sharpe2004} and C$_4$H$_{10}$ (butane)\cite{Sharpe2004}. We initially compute forward models at $R=10\,000$ from 5-12~$\mu$m using line-by-line opacity sampling before binning to the resolution of the observed spectrum to compute the likelihood function. We initialize \texttt{MultiNest} with $2\,000$ live points.

\begin{figure*}
    \centering
    \includegraphics[width=\linewidth]{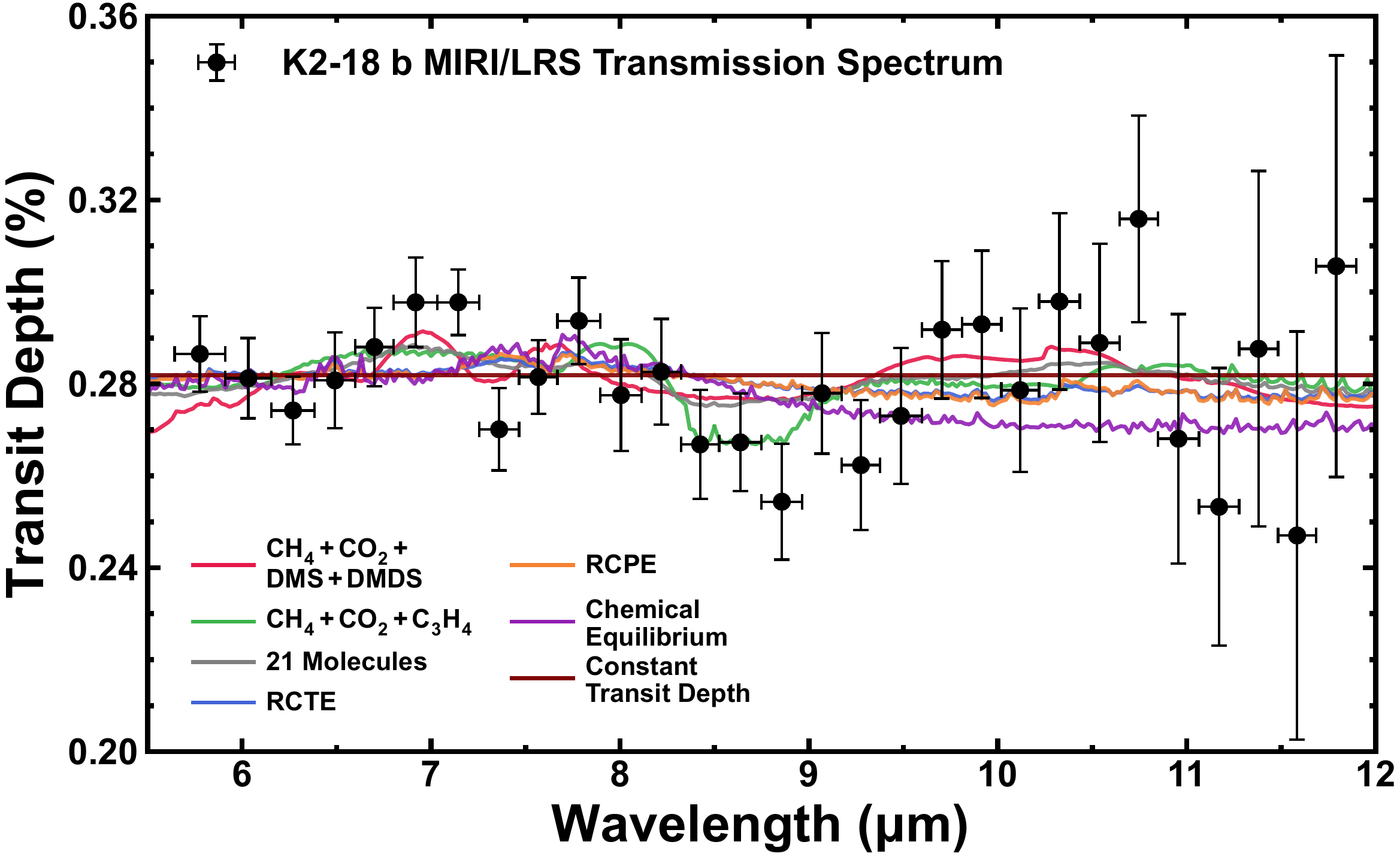}
   \caption{\textbf{Median model fits for the \texttt{JExoRES} reduction of the K2-18 b transmission spectrum.} Median model evaluations for selected atmospheric models applied to the \texttt{JExoRES} dataset. Vertical error bars show $1\sigma$ standard deviations on the measured transit depth; horizontal bars indicate wavelength bin widths. Each point represents a single spectral bin. The Anderson–Darling and Kolmogorov–Smirnov tests find the residuals statistically consistent with a Normal distribution, indicating that the data cannot distinguish between these models. Equivalent fits are obtained for the \texttt{JexoPipe} reduction (not shown).}
    \label{fig:ED4}
\end{figure*}

As with previous modeling approaches, the retrieved posteriors from this parametric model approach are largely unconstrained across both data reductions, that is, the resulting posteriors largely reflect their priors. The only exception is 
propyne (C$_3$H$_4$) which, in the case of the \texttt{JExoRES} reduction, exhibits a long tail towards the lower end of its prior ($10^{-20}$) but has a prominent peak near an abundance of $10^{-3}$, in line with the predicted abundances for K2-18b\cite{Huang2024}. The resulting posterior is not Gaussian with a median and 68.26-percentile range of $\log_{10}\rm{(X_{C_3H_4})}=-6^{+4}_{-8}$. This model provides a good fit to the data as shown in Figure~\ref{fig:spectrum} and discussed below. We further perform additional retrievals removing one gas at a time to assess the model preference for this larger model including all species (the reference model) relative to the simpler models without an absorber. Overall, no individual gas provides any significant change to the model evidence. Only 5 (2) individual gases result in a model preference in support of their inclusion in the reference model for the \texttt{JExoRES} (\texttt{JexoPipe}) reduction, as shown in Extended Data Table \ref{tab:big_retrieval}, when considering the error in the estimated Bayes factors $\Delta(\ln(\mathcal{B}))=0.07$. We proceed to consider two additional nested models; one in which both DMS and DMDS are removed and another where we remove C$_3$H$_4$, C$_4$H$_{8}$ and C$_4$H$_{10}$. Compared to the reference model, these nested models without two and three species respectively result in a $<2\sigma$ decrease in model evidence in favor of the reference model. Extended Data Table \ref{tab:big_retrieval}, shows the relative preference of the full model (reference model) relative to the nested models we tested, and while none meet a $2\sigma$-threshold, the top 3 combinations are DMS+DMDS, C$_3$H$_4$, and the three gas combination of C$_3$H$_4$+C$_4$H$_{8}$+C$_4$H$_{10}$.

After the initial preparation of this manuscript, subsequent work has suggested additional candidate molecules which could be present in the atmosphere of K2-18~b\citeApp{Pica-Ciamarra2025,Hu2025}. We extend this analysis by expanding the model to include 29 absorbers, incorporating all species shown in Figure~\ref{fig:xsec}, along with additional molecules motivated by these concurrent analyses. These include diethyl sulfide ((CH$_3$CH$_2$)$_2$S) and methyl acrylonitrile (C$_4$H$_5$N), proposed as candidate absorbers for the MIRI spectrum\citeApp{Pica-Ciamarra2025}, and methyl mercaptan (CH$_3$SH), reported to show marginal signals across multiple instruments\citeApp{Hu2025}. When all 29 species are included as free parameters, no molecule yields an improvement in Bayesian evidence in both the \texttt{JExoRES} and \texttt{JexoPipe} reductions. Only four species in \texttt{JExoRES} and two in \texttt{JexoPipe} show marginal improvements in model evidence ($\mathcal{B} \leq 1.3 \pm 1.0$), corresponding to less than 2$\sigma$ preference in all cases. Just like DMS and DMDS, the proposed candidates diethyl sulfide and methyl acrylonitrile are not preferred by either MIRI data reduction when comparing this extended 29 chemical species model to the corresponding nested models.

\subsection*{Assessment of Model Fits}\label{sec:model_fits}

We analyze the residuals to the \texttt{JExoRES} data for the self-consistent 1D-RCTE and 1D-RCPE models, chemical equilibrium model, and the parametric models (i.e., free-retrieval models) for CH$_4$, CO$_2$, DMS, and DMDS; CH$_4$, CO$_2$, and C$_3$H$_4$; and the model with 21 molecules. These residuals are shown in Figure \ref{fig:spectrum}, alongside those for the constant transit depth already described above, and were interpreted using the KS and AD tests also described in the section above. We find that all model residuals are consistent with the white noise assumed in the modeling which suggests that all models considered are sufficient to explain the data. Specifically, we find KS test values (and corresponding $p$-values for the hypothesis of no data-model residuals inconsistent with white Gaussian noise) with respect to the median spectrum (obtained from posterior model realizations) of 0.08 (0.99), 0.100 (0.90), 0.17 (0.35), 0.14 (0.60), 0.24 (0.05), and  0.08 (0.99), for the \texttt{ScCHIMERA} 1D-RCTE, \texttt{ScCHIMERA} 1D-RCPE, chemical equilibrium with \texttt{PICASO} and the three parametric models (with DMS and DMDS, propyne, and 21 molecules), respectively. For the same models, the AD test yields 0.157 (0.95), 0.14 (0.98), 0.22 (0.83), 0.29 (0.62), 0.27 (0.68), and 0.20 (0.88), respectively. The median models are shown in Extended Data Figure \ref{fig:ED4}.

\section*{Simplified Models for K2-18~b}\label{sec:k218b-reproduction}

We explore the possibility for simplified models, with fewer free parameters, to explain the MIRI observations of K2-18~b. For these models, we use the \textsc{Aura} atmospheric modeling code\citeApp{Pinhas2018a, Welbanks2019a,Welbanks2022,Nixon2020a,Nixon2022} employed in ref.\cite{Madhusudhan2025}. All parameter choices and priors follow those described in ref.\cite{Madhusudhan2025} where available. We use a pressure grid consisting of 100 uniform layers in log space between 10$^{-6}$ and 10 bar, matching the prior range for $P_{c}$ from ref.\cite{Madhusudhan2025}. We use the same cross-section data as in our larger \textit{Aurora} model. Forward models are computed at $R=10\,000$ before binning. We verified that using higher-resolution forward models ($R=20\,000$) make a negligible difference to our results.

\begin{table*}[ht]
\centering
\caption{ \justifying \textbf{Bayesian model comparison for dimethyl sulfide (DMS) and dimethyl disulfide (DMDS) in the atmosphere of K2-18 b using evidences from this study for a model with 4 gases.} Comparison of models evaluating the relative preference (reported as ``detection significance") for DMS and/or DMDS, using $\ln \mathcal{Z}$ values calculated in this work from the ``canonical" model construction of ref.\cite{Madhusudhan2025}. Each comparison is made relative to a reference model containing CH$_4$, CO$_2$, DMS, and DMDS. Reported values correspond to the logarithmic Bayes factors ($\ln \mathcal{B}$) for each case. Results are provided for the \texttt{JExoRES} and \texttt{JexoPipe} reductions.}
\begin{tabular}{llccc}
\hline
\textbf{Pipeline} & \textbf{Model - This work} & $\boldsymbol{\ln(\mathcal{Z})}$ & $\boldsymbol{\ln(\mathcal{B})}$ & \textbf{Model preference} \\
\hline
\multirow{4}{*}{\texttt{JExoRES}} 
& CH$_4$, CO$_2$, DMS, DMDS & 215.37 & Reference & Reference     \\
& Remove DMS                & 214.98 & 0.38      & 1.6$\sigma$   \\
& Remove DMDS               & 215.15 & 0.22      & 1.4$\sigma$   \\
& Remove DMS and DMDS       & 212.28 & 3.08      & 3.0$\sigma$   \\
\hline
\multirow{4}{*}{\texttt{JexoPipe}} 
& CH$_4$, CO$_2$, DMS, DMDS & 210.95 & Reference & Reference     \\
& Remove DMS                & 210.24 & 0.71      & 1.8$\sigma$   \\
& Remove DMDS               & 210.84 & 0.11      & 1.2$\sigma$   \\
& Remove DMS and DMDS       & 207.77 & 3.18      & 3.0$\sigma$   \\
\hline
\end{tabular}
\label{tab:EDTable2}
\end{table*}

\subsection*{Reconstruction of previous work}\label{sec:reproduction}

We reconstruct the ``canonical'' model constructed by ref.\cite{Madhusudhan2025}. Following their study, we include 16 free parameters: 6 to describe the $P$--$T$ profile, 4 to describe clouds and hazes, $P_{\rm ref}$, a free offset, and 4 molecules (CH$_4$, CO$_2$, DMS and DMDS). We adopt the same priors on all parameters following their Table 3. In the event where a particular model or parameter choice is unclear from the manuscript, we adopt the same choices used in our \textit{Aurora} model. We explore a series of 4 models: one with all 4 species (chosen as the reference model), one with DMS removed, one with DMDS removed, and one with both DMS and DMDS removed. We use relative Bayesian evidences to compute the model preference for DMS, DMDS and the combination of DMDS, as shown in Extended Data Table \ref{tab:EDTable2}. Modeling the \texttt{JExoRES} spectrum, we find a 1.6$\sigma$ preference for the reference model over the model with DMS removed, a 1.4$\sigma$ preference over the model with DMDS removed, and a 3.0$\sigma$ preference over the model with both DMS and DMDS removed. For the \texttt{JexoPipe} spectrum, we find a 1.8$\sigma$ preference for the reference model over the model with DMS removed, a 1.2$\sigma$ preference over the model with DMDS removed, and a 3.0$\sigma$ preference over the model with both DMS and DMDS removed. We find a similar preference for the combination of DMS and DMDS as ref.\cite{Madhusudhan2025}. We discuss the likely origin for the differences in reported significances below. The reconstruction of the ``canonical model'' is shown in Extended Data Figure \ref{fig:ED4}, where the median model realization, from the resulting posterior samples of our parameter estimation, is shown.

\begin{table*}[ht]
\centering
\caption{\justifying \textbf{Bayesian model comparison for dimethyl sulfide (DMS) and dimethyl disulfide (DMDS) using published evidences for a model with 4 gases.} Model preferences (``detection significances") for DMS and/or DMDS in the atmosphere of K2-18 b, adopting $\ln \mathcal{Z}$ values from ref.\cite{Madhusudhan2025}. Comparisons are made relative to a common reference model containing CH$_4$, CO$_2$, DMS, and DMDS. Results are provided for the \texttt{JExoRES} and \texttt{JexoPipe} reductions.}
\begin{tabular}{llccc}
\hline
\textbf{Pipeline} & \textbf{Model - ref\cite{Madhusudhan2025}} & $\boldsymbol{\ln(\mathcal{Z})}$ & $\boldsymbol{\ln(\mathcal{B})}$ & \textbf{Model preference} \\
\hline
\multirow{4}{*}{\texttt{JExoRES}} 
& CH$_4$, CO$_2$, DMS, DMDS & 216.40 & Reference & Reference \\
& Remove DMS                 & 216.40 & 0.00      & N/A  \\
& Remove DMDS                & 215.45 & 0.95      & 2.0$\sigma$ \\
& Remove DMS and DMDS       & 212.59 & 3.80      & 3.2$\sigma$ \\
\hline
\multirow{4}{*}{\texttt{JexoPipe}} 
& CH$_4$, CO$_2$, DMS, DMDS  & 211.59 & Reference & Reference \\
& Remove DMS                    & 211.18 & 0.41      & 1.6$\sigma$ \\
& Remove DMDS                   & 211.18 & 0.41      & 1.6$\sigma$ \\
& Remove DMS and DMDS            & 208.00 & 3.59      & 3.2$\sigma$ \\
\hline
\end{tabular}
\label{tab:EDTable3}
\end{table*}

\subsection*{Interpretation According to Conventional Practices}\label{sec:model_comparison}

The common practice in the field of exoplanetary atmospheric studies is to evaluate the significance of a chemical detection by comparing the Bayesian evidence of a large model, including the chemical of interest, to a nested model in which that species is removed. This approach, formalized in ref.\cite{Benneke2013}, established that detection significances are measured relative to a model covering the full prior hypothesis space rather than relative to a minimal or featureless model. Previous studies\cite{ Welbanks2019b,Madhusudhan2020, Madhusudhan2023} exemplify the wide adoption of this convention.

In ref.\cite{Madhusudhan2025}, the canonical atmospheric model for K2-18~b includes four gases (CH$_4$, CO$_2$, DMS, and DMDS) along with  parameters describing the pressure–temperature profile and cloud properties, for a total of 16 free parameters. Following the prevalent practice, the significance of detecting DMS or DMDS individually would require comparing the Bayesian evidence of the four-gas model against that of models where one gas at a time is removed. Specifically:

\begin{itemize}
    \item Detection of DMS: canonical model vs. model with CH$_4$, CO$_2$, and DMDS.
    \item Detection of DMDS: canonical model vs. model with CH$_4$, CO$_2$, and DMS.
    \item Detection of DMS and DMDS together: canonical model vs. model with only CH$_4$ and CO$_2$.
\end{itemize}

However, inspection of Table 2 in ref.\cite{Madhusudhan2025} suggests that the reported detection significances are relative to the CH$_4$+CO$_2$-only model, not to the full four-gas canonical model. Subtracting the reported $\ln(\mathcal{B})$ values from the $\ln(\mathcal{Z})$ values reveals that the reference model used is the CH$_4$+CO$_2$-only model.

In Extended Data Table~\ref{tab:EDTable3} we present the model preferences, and the associated detection significances of DMS, DMDS, and DMS+DMDS, when the four molecule model is used as reference following the convention in ref.\cite{Benneke2013} instead of using the simpler CH$_4$+CO$_2$-only model as reference as in ref.\cite{Madhusudhan2025}.

Under the conventional approach of using the canonical model as the reference, there is no significant detection of DMS or DMDS individually ($<2\sigma$), and a $\sim$3.2$\sigma$ preference for the combined presence of DMS and DMDS. In contrast, when using the minimal CH$_4$+CO$_2$ model as the reference, the reported detection significances are enhanced. 

\begin{figure}
    \centering
    \includegraphics[width=\linewidth]{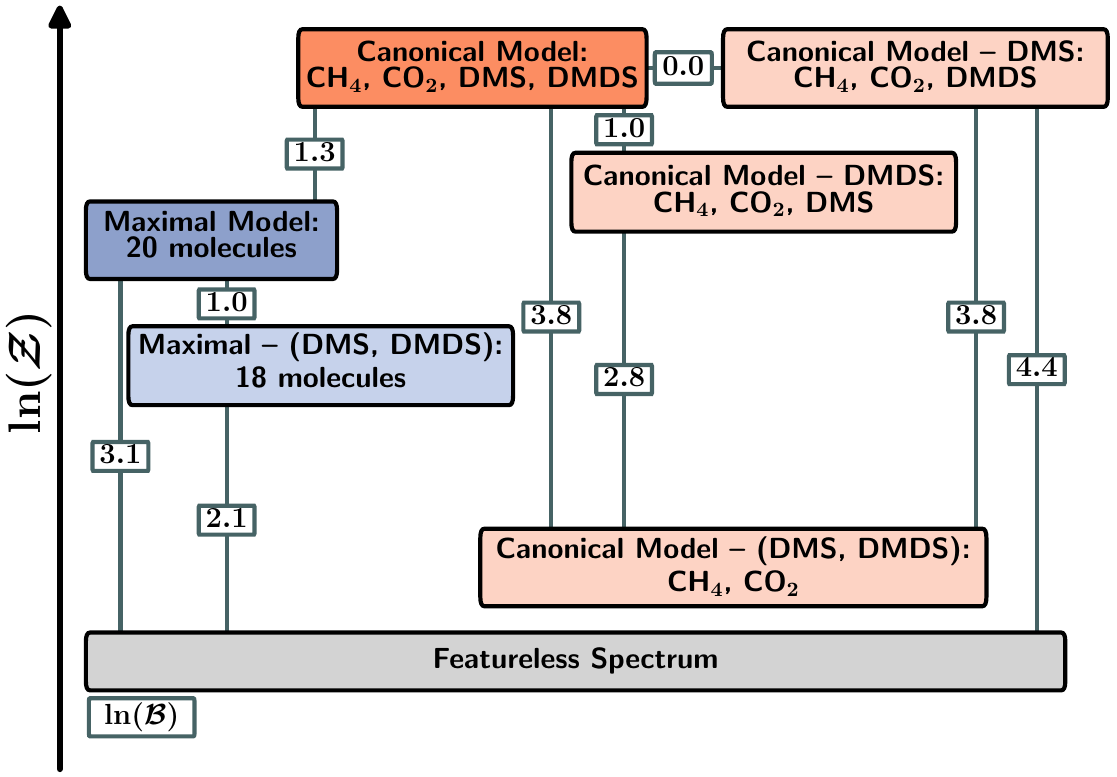}
    \caption{\textbf{Schematic representation of model comparisons and relative Bayesian evidence.} Diagram showing the models listed in Extended Data Table~\ref{tab:EDTable3}, connected by green lines representing the absolute difference in Bayesian evidence ($\vert \ln(B) \vert$) between model pairs. Each link illustrates how the apparent detection significance depends on the specific pair of models compared. Lines represent calculated differences in Bayesian evidence between models.}
    \label{fig:ED5}
\end{figure}

We illustrate the relative nature of Bayesian model comparisons in Extended Data Figure~\ref{fig:ED5}, using the structure introduced in Figure~\ref{fig:model_space}. Each colored box represents a specific model, vertically ordered by increasing Bayesian evidence ($\ln \mathcal{Z}$). The connecting lines indicate pairwise model comparisons, with the associated Bayes factors ($\ln \mathcal{B}$) shown as labels. Some of the Bayes factors shown are directly computed, while others are inferred from reported $\sigma$-values in ref.\cite{Madhusudhan2025} and are included here for illustrative purposes only; their exact values are not reported in ref.\cite{Madhusudhan2025} and may differ slightly from those shown here.

The orange boxes in Extended Data Figure~\ref{fig:ED5} correspond to the family of ``canonical models" reconstructed from ref.\cite{Madhusudhan2025} (see also Extended Data Table~\ref{tab:EDTable3}). The gray and blue boxes represent, respectively, a flat-line (featureless) model and the maximal 20-species model and its nested variant excluding DMS and DMDS. The reported model preference for DMS and DMDS depends on which pair of models is being compared: for instance, comparing the canonical model to a minimal CH$_4$+CO$_2$ baseline yields a stronger preference than comparing against the 18-species nested maximal model. Likewise, whether one uses the best-performing baseline or a worse-performing model as reference strongly influences the resulting Bayes factor.

This case study highlights a broader principle central to this work: Bayesian model comparison quantifies the relative adequacy of models, but does not, by itself, establish the detection of a specific atmospheric constituent. When detection significances are reported without specification of the reference model and the extent of the model space explored, they risk conveying an unwarranted sense of certainty. The interpretation of any apparent detection must therefore acknowledge the conditional nature of model comparisons. Without these caveats, apparent ``detections" of gases can arise from procedural artifacts rather than robust evidence, underscoring the fallacy at the heart of current atmospheric inference practices.

The pitfalls of neglecting the relative nature of Bayesian model comparisons are further exemplified in ref.\cite{Pica-Ciamarra2025}, submitted while this manuscript was under review. Their search for potential trace species in the atmosphere of K2-18~b is also based entirely on comparisons between CH$_4$+CO$_2$+X models and a CH$_4$+CO$_2$-only baseline, rather than comparing nested models within a comprehensive hypothesis space. As we show in Extended Data Figure \ref{fig:ED5}, this methodology can inflate model preferences for certain species depending on the limited subset of the model space under consideration. Similarly to the reported preferences in ref.\cite{Madhusudhan2025}, the $\sim$3$\sigma$ preferences for DMS, diethyl sulfide, and methyl acrylonitrile presented in ref.\cite{Pica-Ciamarra2025} are not robust detections; they are artifacts of a constrained comparison strategy. When evaluated within a larger 29-species hypothesis space (see previous section), none of these molecules results in a meaningful model improvement in either the \texttt{JExoRES} or \texttt{JexoPipe} MIRI spectra of K2-18~b.

\subsection*{Alternative Simplified Models}\label{sec:hydrocarbon}

We consider a number of additional candidate models that could provide  possible explanations for the composition of the atmosphere. Specifically, we examine the hypothesis that hydrocarbons may be present in the atmosphere of K2-18~b. This is motivated by two factors: (1) the presence of CH$_4$ in the planet's atmosphere, as suggested by its JWST NIRISS+NIRSpec spectra\cite{Madhusudhan2023,Schmidt2025}, and (2) the prevalence of a range of hydrocarbons on solar system objects such as Uranus, Neptune and Titan\citeApp{Lunine1993}$^\text{,}$\cite{Waite2005,Burgdorf2006}. We construct a set of models consisting of CH$_4$+CO$_2$+C$_x$H$_y$ for hydrocarbons with pre-computed cross-sections available in the HITRAN database\cite{Sharpe2004,Gordon2022}, alongside several other species predicted from photochemical models for which full line list data is available (C$_2$H$_2$\citeApp{Rothman2013}, C$_2$H$_4$\citeApp{Mant2018} and C$_2$H$_6$\citeApp{Rothman2013}). Following the approach used in ref.\cite{Madhusudhan2025} for DMS and DMDS, for molecules with pre-computed cross-sections rather than complete line lists, we adopt the cross section at 298~K and 1~bar, although we note that not accounting for pressure- and temperature-dependent effects on the cross-sections introduces further uncertainty into our analysis. We also caution that, depending on the pressures and temperatures being probed in these transmission observations, some of the species considered here may not be expected to be present in gaseous form. In total, we consider 90 molecules, and conduct a set of 90 retrievals, comparing each one to a model with only CH$_4$+CO$_2$ to measure model preference in the absence of any other chemical species. In these cases, we adopt the same priors as ref.\cite{Madhusudhan2025}.

Our results are shown in Extended Data Tables \ref{tab:EDTable4} and \ref{tab:EDTable5}. Fitting to the \texttt{JExoRES} spectrum, we obtain $\ln \mathcal{Z} = 212.28$ for the CH$_4$+CO$_2$-only model, and we find that every molecule we consider increases the Bayesian evidence to some degree. In 59/90 cases, the model including the chosen hydrocarbon is preferred to the model without it at $\geq2\sigma$. We find 6 species whose level of preference is greater than that of DMDS: propyne (3.09$\sigma$), cyclohexane (3.03$\sigma$), 2-butene (2.99$\sigma$), 2-pentene (2.92$\sigma$), butane (2.92$\sigma$) and cyclopentane (2.85$\sigma$). Of these, all except cyclopentane achieve a greater model preference than DMS, and propyne and cyclohexane each individually achieve a greater model preference than the combined inference of DMS and DMDS. The median model from the retrievals including propyne is shown in Extended Data Figure \ref{fig:ED4}. For the \texttt{JexoPipe} spectrum, we find 36/90 hydrocarbons to be preferred at $\geq2\sigma$, with 3 whose level of preference exceeds that of DMDS: cyclohexane (3.04$\sigma$), propyne (3.02$\sigma$) and butane (2.84$\sigma$). Cyclohexane and propyne individually show a greater model preference than the combination of DMS and DMDS.

By combining different hydrocarbons, one can further increase the Bayesian evidence, creating an even higher model preference relative to the base model. We also find that including H$_2$O in our model typically increases the evidence. We do not consider all possible permutations due to the computational cost: even iterating through all pairs of hydrocarbons would require 4005 retrievals. However, we do consider some pairings involving the species that yielded the largest evidence increases. In some cases, we constructed models such that including DMS and DMDS led to a \textit{decrease} in Bayesian evidence: for example, a model consisting of CH$_4$, CO$_2$, H$_2$O, propyne and 2-methylpropane has $\ln \mathcal{Z}=215.83$ when fit to the \texttt{JExoRES} spectrum, whereas CH$_4$, CO$_2$, H$_2$O, DMS, DMDS, propyne and 2-methylpropane has $\ln \mathcal{Z}=215.69$, leading to a 1.29$\sigma$ preference for the \textit{absence} of DMS and DMDS. We conclude that one can only find a ``detection'' of DMS or DMDS based on these observations by excluding hydrocarbon species from one's model. If hydrocarbons such as propyne are permitted in the model, then there is no longer any evidence for DMS or DMDS.

As noted in the main text, this does not mean we have ``detected'' hydrocarbons in the atmosphere of K2-18~b from its JWST/MIRI transmission spectrum. Indeed, for the hydrocarbon species included in our larger retrieval model with 21 gases, none of them revealed a high model preference. For example, propyne (C$_3$H$_4$) is preferred in the \texttt{JExoRES} spectrum at 3.1$\sigma$ significance when compared against a CO$_2$+CH$_4$-only model, but at just 1.5$\sigma$ significance when compared against the larger model. Each of the above detection significances must be interpreted in the context of the model being applied. In practice, the construction of a fully exhaustive model remains out of reach, but we must strive to at least consider reasonable alternative hypotheses before making strong claims.

\begin{table*}[ht]
\centering
\caption{\justifying \textbf{Bayesian model comparison for hydrocarbon candidates in the atmosphere of K2-18 b (part 1 of 2).} Model preferences for named hydrocarbons, expressed as the difference in Bayesian evidence relative to a model including only CH$_4$ and CO$_2$. For \texttt{JExoRES} (\textdagger), the reference model has $\ln \mathcal{Z} = 212.28$ ; for \texttt{JexoPipe} (\textdaggerdbl), $\ln \mathcal{Z} = 207.77$. Reported ``detection significances" correspond to the relative preference for models including CH$_4$ and CO$_2$, and one additional species. For comparison, DMS and DMDS yield preferences of 2.90$\sigma$ (DMS), 2.83$\sigma$ (DMDS), and 2.98$\sigma$ (DMS+DMDS) for \texttt{JExoRES} and 2.98$\sigma$ (DMS), 2.74$\sigma$ (DMDS) and 3.02$\sigma$ (DMS+DMDS) for \texttt{JexoPipe}.}
\resizebox{\textwidth}{!}{%
\begin{tabular}{l|cc|cc|cc|cc}
\hline
\textbf{Molecule (Formula)} 
& \multicolumn{2}{c|}{$\boldsymbol{\ln(\mathcal{Z})}$} 
& \multicolumn{2}{c|}{$\boldsymbol{\chi^2 / N_{\mathrm{data}}}$} 
& \multicolumn{2}{c|}{$\boldsymbol{\ln(\mathcal{L}_{\max})}$} 
& \multicolumn{2}{c}{\textbf{Preference} ($\boldsymbol{\sigma}$)} \\
& \textdagger & \textdaggerdbl & \textdagger & \textdaggerdbl & \textdagger & \textdaggerdbl & \textdagger & \textdaggerdbl \\
\hline
Propyne (C$_3$H$_4$)  & 215.67 & 210.96 & 0.68 & 0.72 & 220.08 & 215.16 & 3.09 & 3.02 \\
Cyclohexane (C$_6$H$_{12}$)  & 215.49 & 211.01 & 0.68 & 0.68 & 220.17 & 215.66 & 3.03 & 3.04 \\
2-Butene (C$_4$H$_8$)  & 215.40 & 209.14 & 0.62 & 0.76 & 221.09 & 214.56 & 2.99 & 2.24 \\
trans-2-Pentene (C$_5$H$_{10}$)  & 215.22 & 209.30 & 0.63 & 0.80 & 220.92 & 214.04 & 2.92 & 2.32 \\
Butane (C$_4$H$_{10}$)  & 215.20 & 210.49 & 0.71 & 0.74 & 219.68 & 214.85 & 2.92 & 2.84 \\
Cyclopentane (C$_5$H$_{10}$)  & 215.02 & 210.09 & 0.67 & 0.76 & 220.25 & 214.63 & 2.85 & 2.68 \\
n-Tetradecane (C$_{14}$H$_{30}$)  & 214.51 & 210.20 & 0.78 & 0.76 & 218.67 & 214.58 & 2.64 & 2.73 \\
Isobutene (C$_4$H$_8$)  & 214.27 & 209.44 & 0.76 & 0.81 & 219.03 & 213.91 & 2.54 & 2.39 \\
cis-2-Pentene (C$_5$H$_{10}$)  & 214.19 & 208.93 & 0.74 & 0.78 & 219.26 & 214.32 & 2.50 & 2.12 \\
Propene (C$_3$H$_6$)  & 214.14 & 208.50 & 0.75 & 0.81 & 219.19 & 213.89 & 2.48 & 1.85 \\
2,2-Dimethylbutane (C$_6$H$_{14}$)  & 214.04 & 208.88 & 0.82 & 0.85 & 218.05 & 213.23 & 2.43 & 2.09 \\
Ethylene (C$_2$H$_4$)  & 214.03 & 208.55 & 0.82 & 0.86 & 218.14 & 213.07 & 2.43 & 1.88 \\
1-Pentene (C$_5$H$_{10}$)  & 214.01 & 209.09 & 0.77 & 0.76 & 218.85 & 214.51 & 2.42 & 2.21 \\
Cycloheptane (C$_7$H$_{14}$)  & 213.99 & 208.76 & 0.82 & 0.88 & 218.18 & 212.85 & 2.41 & 2.02 \\
Cyclopropane (C$_3$H$_6$)  & 213.95 & 208.36 & 0.75 & 0.87 & 219.15 & 213.02 & 2.39 & 1.74 \\
1-Heptene (C$_7$H$_{14}$)  & 213.91 & 209.10 & 0.78 & 0.78 & 218.67 & 214.32 & 2.37 & 2.22 \\
2-Methyl-1-pentene (C$_6$H$_{12}$)  & 213.87 & 209.06 & 0.83 & 0.83 & 218.00 & 213.52 & 2.35 & 2.19 \\
Ethane (C$_2$H$_6$)  & 213.87 & 208.73 & 0.82 & 0.86 & 218.10 & 213.06 & 2.34 & 2.00 \\
3-Methylpentane (C$_6$H$_{14}$)  & 213.85 & 209.16 & 0.82 & 0.82 & 218.15 & 213.73 & 2.33 & 2.24 \\
1-Butene (C$_4$H$_8$)  & 213.84 & 208.79 & 0.81 & 0.78 & 218.23 & 214.33 & 2.33 & 2.04 \\
4-Methyl-2-pentene (C$_6$H$_{12}$)  & 213.82 & 208.67 & 0.83 & 0.85 & 217.93 & 213.31 & 2.32 & 1.96 \\
Isopentane (C$_5$H$_{12}$)  & 213.82 & 209.71 & 0.85 & 0.77 & 217.67 & 214.43 & 2.32 & 2.51 \\
2-Methyl-1-butene (C$_5$H$_{10}$)  & 213.80 & 208.78 & 0.82 & 0.84 & 218.12 & 213.39 & 2.31 & 2.03 \\
Styrene (C$_8$H$_8$)  & 213.80 & 208.53 & 0.78 & 0.83 & 218.69 & 213.53 & 2.31 & 1.87 \\
n-Heptane (C$_7$H$_{16}$)  & 213.75 & 209.02 & 0.83 & 0.85 & 218.02 & 213.25 & 2.28 & 2.17 \\
1-Hexene (C$_6$H$_{12}$)  & 213.73 & 208.91 & 0.77 & 0.70 & 218.78 & 215.47 & 2.27 & 2.11 \\
3-Methylhexane (C$_7$H$_{16}$)  & 213.72 & 208.98 & 0.84 & 0.83 & 217.85 & 213.55 & 2.27 & 2.15 \\
Pentane (C$_5$H$_{12}$)  & 213.69 & 209.36 & 0.85 & 0.80 & 217.72 & 214.04 & 2.26 & 2.35 \\
n-Hexane (C$_6$H$_{14}$)  & 213.63 & 209.14 & 0.87 & 0.83 & 217.46 & 213.57 & 2.22 & 2.23 \\
1-Octene (C$_8$H$_{16}$)  & 213.58 & 208.90 & 0.84 & 0.79 & 217.77 & 214.09 & 2.20 & 2.10 \\
Benzene (C$_6$H$_6$)  & 213.57 & 208.05 & 0.85 & 0.94 & 217.63 & 211.89 & 2.19 & 1.47 \\
1,3-Butadiene (C$_4$H$_6$)  & 213.51 & 208.31 & 0.78 & 0.87 & 218.69 & 213.03 & 2.15 & 1.70 \\
1-Undecene (C$_{11}$H$_{22}$)  & 213.51 & 208.80 & 0.82 & 0.82 & 218.10 & 213.64 & 2.15 & 2.04 \\
Vinyl toluene (C$_9$H$_{10}$)  & 213.50 & 208.26 & 0.85 & 0.88 & 217.73 & 212.81 & 2.15 & 1.66 \\
Isoprene (C$_5$H$_8$)  & 213.47 & 208.41 & 0.87 & 0.89 & 217.39 & 212.73 & 2.13 & 1.78 \\
1-Nonene (C$_9$H$_{18}$)  & 213.46 & 208.36 & 0.90 & 0.93 & 216.92 & 212.09 & 2.13 & 1.74 \\
(-)-$\beta$-Pinene (C$_{10}$H$_{16}$)  & 213.46 & 208.93 & 0.84 & 0.80 & 217.83 & 214.06 & 2.13 & 2.12 \\
n-Decane (C$_{10}$H$_{22}$)  & 213.46 & 208.91 & 0.87 & 0.85 & 217.42 & 213.33 & 2.13 & 2.11 \\
4-Methyl-1-pentene (C$_6$H$_{12}$)  & 213.44 & 208.70 & 0.85 & 0.80 & 217.66 & 213.94 & 2.12 & 1.98 \\
Octane (C$_8$H$_{18}$)  & 213.44 & 208.95 & 0.84 & 0.85 & 217.82 & 213.23 & 2.12 & 2.13 \\
1-Decene (C$_{10}$H$_{20}$)  & 213.43 & 208.83 & 0.82 & 0.81 & 218.06 & 213.81 & 2.11 & 2.06 \\
n-Nonane (C$_9$H$_{20}$)  & 213.42 & 208.90 & 0.87 & 0.85 & 217.36 & 213.24 & 2.10 & 2.10 \\
1,2,3-Trimethylbenzene (C$_9$H$_{12}$)  & 213.38 & 208.19 & 0.89 & 0.88 & 217.05 & 212.76 & 2.08 & 1.60 \\
2,4,4-Trimethyl-1-pentene (C$_8$H$_{16}$)  & 213.38 & 208.68 & 0.89 & 0.86 & 217.14 & 213.08 & 2.08 & 1.97 \\
sec-Butylbenzene (C$_{10}$H$_{14}$)  & 213.38 & 208.74 & 0.89 & 0.86 & 217.03 & 213.18 & 2.08 & 2.00 \\
\hline
\end{tabular}
}
\label{tab:EDTable4}
\end{table*}

\begin{table*}[ht]
\centering
\caption{\justifying \textbf{ Bayesian model comparison for hydrocarbon candidates (part 2 of 2).} Continuation of Extended Data Table \ref{tab:EDTable4}. Model preferences are ranked by Bayesian evidence ($\ln \mathcal{Z}$) relative to the  CH$_4$ and CO$_2$ reference model. Reported significances denote the relative preference for each hydrocarbon species.}
\resizebox{\textwidth}{!}{%
\begin{tabular}{l|cc|cc|cc|cc}
\hline
\textbf{Molecule (Formula)} 
& \multicolumn{2}{c|}{$\boldsymbol{\ln(\mathcal{Z})}$} 
& \multicolumn{2}{c|}{$\boldsymbol{\chi^2 / N_{\mathrm{data}}}$} 
& \multicolumn{2}{c|}{$\boldsymbol{\ln(\mathcal{L}_{\max})}$} 
& \multicolumn{2}{c}{\textbf{Preference} ($\boldsymbol{\sigma}$)} \\
& \textdagger & \textdaggerdbl & \textdagger & \textdaggerdbl & \textdagger & \textdaggerdbl & \textdagger & \textdaggerdbl \\
\hline
Isooctane (C$_8$H$_{18}$)  & 213.37 & 208.71 & 0.91 & 0.91 & 216.81 & 212.42 & 2.07 & 1.99 \\
3-Carene (C$_{10}$H$_{16}$)  & 213.34 & 208.75 & 0.89 & 0.84 & 217.06 & 213.38 & 2.06 & 2.01 \\
Cyclodecane (C$_{10}$H$_{20}$)  & 213.34 & 208.91 & 0.90 & 0.88 & 217.00 & 212.81 & 2.06 & 2.11 \\
2-Methyl-2-butene (C$_5$H$_{10}$)  & 213.32 & 208.17 & 0.88 & 0.88 & 217.18 & 212.78 & 2.04 & 1.59 \\
DL-Limonene (C$_{10}$H$_{16}$)  & 213.31 & 208.31 & 0.88 & 0.86 & 217.21 & 213.10 & 2.04 & 1.70 \\
4-Ethyltoluene (C$_9$H$_{12}$)  & 213.31 & 208.12 & 0.92 & 0.93 & 216.74 & 212.15 & 2.04 & 1.54 \\
n-Undecane (C$_{11}$H$_{24}$)  & 213.30 & 208.93 & 0.90 & 0.85 & 216.91 & 213.22 & 2.03 & 2.12 \\
Isodurene (C$_{10}$H$_{14}$)  & 213.29 & 207.99 & 0.91 & 0.92 & 216.86 & 212.26 & 2.03 & 1.40 \\
Isohexane (C$_6$H$_{14}$)  & 213.28 & 208.48 & 0.92 & 0.90 & 216.73 & 212.47 & 2.02 & 1.83 \\
n-Tridecane (C$_{13}$H$_{28}$)  & 213.28 & 208.69 & 0.90 & 0.88 & 217.00 & 212.80 & 2.02 & 1.98 \\
3-Methyl-1-butene (C$_5$H$_{10}$)  & 213.27 & 209.04 & 0.87 & 0.77 & 217.41 & 214.47 & 2.02 & 2.18 \\
Mesitylene (C$_9$H$_{12}$)  & 213.26 & 207.99 & 0.89 & 0.92 & 217.06 & 212.24 & 2.01 & 1.39 \\
1,2,3,4-Tetramethylbenzene (C$_{10}$H$_{14}$)  & 213.24 & 208.17 & 0.90 & 0.90 & 217.02 & 212.49 & 2.00 & 1.58 \\
Ethylbenzene (C$_8$H$_{10}$)  & 213.24 & 208.20 & 0.92 & 0.91 & 216.62 & 212.43 & 2.00 & 1.61 \\
2-Methyl-2-pentene (C$_6$H$_{12}$)  & 213.21 & 208.41 & 0.88 & 0.87 & 217.21 & 212.92 & 1.98 & 1.78 \\
tert-Butylbenzene (C$_{10}$H$_{14}$)  & 213.16 & 208.40 & 0.92 & 0.92 & 216.65 & 212.19 & 1.95 & 1.77 \\
D-Limonene (C$_{10}$H$_{16}$)  & 213.16 & 208.47 & 0.93 & 0.87 & 216.50 & 212.91 & 1.95 & 1.83 \\
2,3-Dimethylbutane (C$_6$H$_{14}$)  & 213.15 & 208.55 & 0.93 & 0.92 & 216.45 & 212.30 & 1.94 & 1.88 \\
2-Ethyltoluene (C$_9$H$_{12}$)  & 213.10 & 208.06 & 0.94 & 0.95 & 216.37 & 211.83 & 1.91 & 1.48 \\
2,4,4-Trimethyl-2-pentene (C$_8$H$_{16}$)  & 213.09 & 208.12 & 0.93 & 0.94 & 216.49 & 211.94 & 1.90 & 1.54 \\
Cycloheptene (C$_7$H$_{12}$)  & 213.05 & 208.11 & 0.93 & 0.95 & 216.46 & 211.76 & 1.88 & 1.52 \\
o-Xylene (C$_8$H$_{10}$)  & 213.02 & 207.92 & 0.95 & 0.93 & 216.25 & 212.08 & 1.85 & 1.31 \\
Myrcene (C$_{10}$H$_{16}$)  & 213.02 & 208.38 & 0.95 & 0.86 & 216.29 & 213.17 & 1.85 & 1.76 \\
Isobutane (C$_4$H$_{10}$)  & 213.01 & 208.40 & 0.93 & 0.91 & 216.58 & 212.37 & 1.85 & 1.77 \\
L-$\alpha$-pinene (C$_{10}$H$_{16}$)  & 213.01 & 208.17 & 0.95 & 0.92 & 216.23 & 212.18 & 1.84 & 1.58 \\
Isocumene (C$_9$H$_{12}$)  & 212.99 & 208.12 & 0.90 & 0.94 & 217.02 & 211.93 & 1.83 & 1.54 \\
Propane (C$_3$H$_8$)  & 212.98 & 208.28 & 0.88 & 0.87 & 217.30 & 213.01 & 1.82 & 1.68 \\
1-Butyne (C$_4$H$_6$)  & 212.96 & 209.05 & 0.86 & 0.85 & 217.55 & 213.25 & 1.81 & 2.18 \\
2,4-Dimethylpentane (C$_7$H$_{16}$)  & 212.96 & 208.29 & 0.95 & 0.91 & 216.25 & 212.39 & 1.81 & 1.69 \\
p-Xylene (C$_8$H$_{10}$)  & 212.92 & 208.45 & 0.91 & 0.92 & 216.79 & 212.32 & 1.78 & 1.81 \\
n-Pentadecane (C$_{15}$H$_{32}$)  & 212.92 & 207.95 & 0.96 & 0.96 & 216.14 & 211.61 & 1.78 & 1.35 \\
Cumene (C$_9$H$_{12}$)  & 212.91 & 208.32 & 0.94 & 0.94 & 216.41 & 212.00 & 1.77 & 1.71 \\
Toluene (C$_7$H$_8$)  & 212.89 & 207.80 & 0.95 & 0.97 & 216.23 & 211.46 & 1.75 & 1.06 \\
Cyclooctane (C$_8$H$_{16}$)  & 212.88 & 208.07 & 0.94 & 0.94 & 216.41 & 211.95 & 1.75 & 1.49 \\
Tetralin (C$_{10}$H$_{12}$)  & 212.86 & 208.03 & 0.98 & 0.95 & 215.83 & 211.80 & 1.73 & 1.45 \\
m-Xylene (C$_8$H$_{10}$)  & 212.86 & 207.79 & 0.98 & 0.98 & 215.76 & 211.35 & 1.73 & 1.03 \\
Acetylene (C$_2$H$_2$)  & 212.83 & 208.92 & 0.96 & 0.84 & 216.12 & 213.40 & 1.71 & 2.11 \\
2-Carene (C$_{10}$H$_{16}$)  & 212.82 & 208.17 & 0.89 & 0.91 & 217.16 & 212.41 & 1.70 & 1.58 \\
3-Ethyltoluene (C$_9$H$_{12}$)  & 212.81 & 207.90 & 0.97 & 0.98 & 215.91 & 211.35 & 1.69 & 1.28 \\
Allene (C$_3$H$_4$)  & 212.80 & 207.94 & 0.96 & 0.95 & 216.14 & 211.80 & 1.69 & 1.34 \\
n-Hexadecane (C$_{16}$H$_{34}$)  & 212.79 & 208.13 & 0.94 & 0.92 & 216.38 & 212.24 & 1.68 & 1.55 \\
Naphthalene (C$_{10}$H$_8$)  & 212.76 & 207.73 & 0.98 & 0.98 & 215.83 & 211.33 & 1.65 & 0.91 \\
4-Vinylcyclohexene (C$_8$H$_{12}$)  & 212.75 & 208.19 & 0.96 & 0.88 & 216.02 & 212.77 & 1.65 & 1.60 \\
Cyclopentene (C$_5$H$_8$)  & 212.74 & 207.97 & 0.99 & 0.97 & 215.67 & 211.57 & 1.63 & 1.38 \\
Cyclohexene (C$_6$H$_{10}$)  & 212.65 & 208.14 & 0.98 & 0.94 & 215.76 & 211.99 & 1.55 & 1.56 \\
\hline
\end{tabular}}
\label{tab:EDTable5}
\end{table*}

\subsubsection*{Data Availability}

The JWST/MIRI LRS transmission spectrum of K2-18~b, reduced with the \texttt{JExoRES} and \texttt{JexoPipe} pipelines, are available at \url{https://osf.io/gmhw3}.

\subsubsection*{Code Availability}

The self-consistent modeling framework \texttt{ScCHIMERA} is adapted from \texttt{CHIMERA}, which is available at \url{https://github.com/mrline/CHIMERA}. The chemical equilibrium modeling framework \texttt{PICASO} is available at \url{https://github.com/natashabatalha/picaso/}.

\bmhead{Acknowledgments}

L.W. and M.C.N. were supported by the Heising-Simons Foundation through a 51 Pegasi b Fellowship. P.M. and M.C.N. were supported under grant JWST-AR 06347 (PI: Nixon). P.M. acknowledges that this work was performed under the auspices of the U.S. Department of Energy by Lawrence Livermore National Laboratory under Contract DE-AC52-07NA27344. Computing support for this work came from the Lawrence Livermore National Laboratory 19th Institutional Computing Grand Challenge program (PI: McGill). The document number is LLNL-JRNL-2005017. L.W., L.T. \& M.R.L. acknowledge support from NASA XRP Grant [80NSSC24K0160] (P.I. Welbanks).  D.Z.S. is supported by an NSF Astronomy and Astrophysics Postdoctoral Fellowship under award AST-2303553. This research award is partially funded by a generous gift of Charles Simonyi to the NSF Division of Astronomical Sciences. The award is made in recognition of significant contributions to Rubin Observatory’s Legacy Survey of Space and Time. S.M. is supported by the Templeton Theory-Experiment (TEX) Cross Training Fellowship from the Templeton foundation. SM also acknowledges use of the \texttt{lux} supercomputer at UC Santa Cruz, funded by NSF MRI grant AST 1828315. A.D.F. acknowledges funding from NASA through the NASA Hubble Fellowship grant HST-HF2-51530.001-A awarded by STScI. L.W., L.S.W., M.R.L., Y.R. acknowledge Research Computing at Arizona State University\citeApp{HPC_ASU23} for providing HPC and storage. M.C.N. acknowledges University of Maryland high-performance computing resources used to conduct research presented in this paper. The authors thank Thomas Greene, Jonathan Lunine, and Aaron Bello-Arufe for helpful comments on the manuscript, as well as the anonymous referees for their insightful feedback.

\bmhead{Author Contribution}

L.W.~led the development of the project and co-leads this publication with M.C.N. as both contributions were fundamental for the completion of this manuscript. L.W. and M.C.N. contributed equally to the study. L.W. performed the atmospheric modeling for the 33 parameter model, and validated the 1D-RCTE and 1D-RCPE models. L.W. contributed to the manuscript, figures, and tables preparation, interpretation of observations, and statistical analysis. M.C.N. led the parametric atmospheric modeling and led the investigation of hydrocarbons as well as the production of relevant line lists and sources of opacity. M.C.N. contributed to the manuscript, figures, and tables preparation, interpretation of observations, and statistical analysis. P.M. contributed to the discussions that shaped and formed this project, contributed to the production of atmospheric models, contributed to the statistical analysis in this manuscript, and provided critical interpretation of the results. P.M. contributed to the manuscript, figures, and tables preparation. L.T. created the grids of self-consistent models, both 1D-RCTE and 1D-RCPE. L.T. contributed to the analysis of the observations and performed part of the parameter estimation (retrievals). L.S.W. contributed to the statistical analysis and the production of the atmospheric models and grid-based methodology. Y.R. performed the GP analysis of the observations. L.T., L.S.W., and Y.R. contributed to the manuscript. A.D.F. contributed to the analysis of the observations, text, and figures. S.M. computed the chemical equilibrium parametric model fits with \texttt{PICASO} and computed 1D chemical kinetic models for K2-18 b's atmosphere using \texttt{photochem}. S.M. contributed to the text, figures, and tables. S.S. helped develop the scope of the project and provided feedback to the manuscript. M.R.L. provided guidance and training to the team. M.R.L. developed the 1D-RCPE and 1D-RCTE methodology and contributed to the statistical analysis. B.B. contributed text and comments to the manuscript.  T.B. contributed to the analysis of the observations and statistical analysis. T.B. contributed to the manuscript, figures, and tables. D.Z.S. contributed to the text and provided feedback on the scope of the project. V.P. provided comments to the manuscript and provided input into the statistical analysis in the project. D.K.S. provided comments to the manuscript.

\bmhead{Competing interests} The authors declare no conflict of interests.

\bibliographystyleApp{sn-standardnature}

\end{document}